\documentclass[preprint2]{emulateapj}
\usepackage{amsmath}
\usepackage{natbib}
\usepackage{graphicx}
\usepackage{epsf}
\usepackage{color}
\input{epsf}
\usepackage{epsfig}
\DeclareGraphicsExtensions{.jpg,.pdf,.png,.eps,.ps}
\graphicspath{{FIGURES/}}

\newcommand{\ltsima}{$\; \buildrel < \over \sim \;$}
\newcommand{\ltsim}{\lower.5ex\hbox{\ltsima}}

\newcommand{\twentythree}{$23^\mathrm{h} 30^\mathrm{m}$}
\newcommand{\five}{$5^\mathrm{h} 30^\mathrm{m}$}
\newcommand{\atszcosm}{A_{\rm tSZ}}
\newcommand{\modelletter}{\Phi_\ell}
\newcommand{\modelnorm}{\Phi_{3000}}
\newcommand{\amplitudeletter}{D_{3000}^}
\newcommand{\hmsun}{h^{-1} \; M_{\odot}}

\begin{document}

\title{Improved constraints on cosmic microwave background secondary anisotropies from the complete 2008 South Pole Telescope data}

\author{
 E.~Shirokoff,\altaffilmark{1} 
 C.~L.~Reichardt,\altaffilmark{1}
 L.~Shaw,\altaffilmark{2}
 M.~Millea,\altaffilmark{3}
 P.~A.~R.~Ade,\altaffilmark{4}
 K.~A.~Aird,\altaffilmark{5}
 B.~A.~Benson,\altaffilmark{1,6,7}
 L.~E.~Bleem,\altaffilmark{6,8}
 J.~E.~Carlstrom,\altaffilmark{6,7,8,9}
 C.~L.~Chang,\altaffilmark{6,7}
 H.~M. Cho, \altaffilmark{10}
 T.~M.~Crawford,\altaffilmark{6,9}
 A.~T.~Crites,\altaffilmark{6,9}
 T.~de~Haan,\altaffilmark{11}
 M.~A.~Dobbs,\altaffilmark{11}
 J.~Dudley,\altaffilmark{11}
 E.~M.~George,\altaffilmark{1}
 N.~W.~Halverson,\altaffilmark{12}
 G.~P.~Holder,\altaffilmark{11}
 W.~L.~Holzapfel,\altaffilmark{1}
 J.~D.~Hrubes,\altaffilmark{5}
 M.~Joy,\altaffilmark{13}
 R.~Keisler,\altaffilmark{6,8}
 L.~Knox,\altaffilmark{3}
 A.~T.~Lee,\altaffilmark{1,14}
 E.~M.~Leitch,\altaffilmark{6,9}
 M.~Lueker,\altaffilmark{15}
 D.~Luong-Van,\altaffilmark{5}
 J.~J.~McMahon,\altaffilmark{6,7,16}
 J.~Mehl,\altaffilmark{6}
 S.~S.~Meyer,\altaffilmark{6,7,8,9}
 J.~J.~Mohr,\altaffilmark{17,18,19}
 T.~E.~Montroy,\altaffilmark{20}
 S.~Padin,\altaffilmark{6,9,15}
 T.~Plagge,\altaffilmark{6,9}
 C.~Pryke,\altaffilmark{6,7,9}
 J.~E.~Ruhl,\altaffilmark{20}
 K.~K.~Schaffer,\altaffilmark{6,7,21}
 H.~G.~Spieler,\altaffilmark{14}
 Z.~Staniszewski,\altaffilmark{20}
 A.~A.~Stark,\altaffilmark{22}
 K.~Story,\altaffilmark{6,8}
 K.~Vanderlinde,\altaffilmark{11}
 J.~D.~Vieira,\altaffilmark{15}
 R.~Williamson\altaffilmark{6,9} and
 O.~Zahn\altaffilmark{23}
}

\altaffiltext{1}{Department of Physics,
University of California, Berkeley, CA 94720}
\altaffiltext{2}{Department of Physics, Yale University, P.O. Box 208210, New Haven,
CT 06520-8120}
\altaffiltext{3}{Department of Physics, 
University of California, One Shields Avenue, Davis, CA 95616}
\altaffiltext{4}{Department of Physics and Astronomy,
Cardiff University, CF24 3YB, UK}
\altaffiltext{5}{University of Chicago,
5640 South Ellis Avenue, Chicago, IL 60637}
\altaffiltext{6}{Kavli Institute for Cosmological Physics,
University of Chicago,
5640 South Ellis Avenue, Chicago, IL 60637}
\altaffiltext{7}{Enrico Fermi Institute,
University of Chicago,
5640 South Ellis Avenue, Chicago, IL 60637}
\altaffiltext{8}{Department of Physics,
University of Chicago,
5640 South Ellis Avenue, Chicago, IL 60637}
\altaffiltext{9}{Department of Astronomy and Astrophysics,
University of Chicago,
5640 South Ellis Avenue, Chicago, IL 60637}
\altaffiltext{10}{NIST Quantum Devices Group, 325 Broadway Mailcode 817.03, Boulder, CO 80305, USA}
\altaffiltext{11}{Department of Physics,
McGill University, 3600 Rue University, 
Montreal, Quebec H3A 2T8, Canada}
\altaffiltext{12}{Department of Astrophysical and Planetary Sciences and Department of Physics,
University of Colorado,
Boulder, CO 80309}
\altaffiltext{13}{Department of Space Science, VP62,
NASA Marshall Space Flight Center,
Huntsville, AL 35812}
\altaffiltext{14}{Physics Division,
Lawrence Berkeley National Laboratory,
Berkeley, CA 94720}
\altaffiltext{15}{California Institute of Technology, MS 249-17, 1216 E. California Blvd., Pasadena, CA 91125, USA}
\altaffiltext{16}{Department of Physics, University of Michigan, 450 Church Street, Ann  
Arbor, MI, 48109}
\altaffiltext{17}{Department of Physics,
Ludwig-Maximilians-Universit\"{a}t,
Scheinerstr.\ 1, 81679 M\"{u}nchen, Germany}
\altaffiltext{18}{Excellence Cluster Universe,
Boltzmannstr.\ 2, 85748 Garching, Germany}
\altaffiltext{19}{Max-Planck-Institut f\"{u}r extraterrestrische Physik,
Giessenbachstr.\ 85748 Garching, Germany}
\altaffiltext{20}{Physics Department, Center for Education and Research in Cosmology 
and Astrophysics, 
Case Western Reserve University,
Cleveland, OH 44106}
\altaffiltext{21}{Liberal Arts Department, 
School of the Art Institute of Chicago, 
112 S Michigan Ave, Chicago, IL 60603}
\altaffiltext{22}{Harvard-Smithsonian Center for Astrophysics,
60 Garden Street, Cambridge, MA 02138}
\altaffiltext{23}{Berkeley Center for Cosmological Physics,
Department of Physics, University of California, and Lawrence Berkeley
National Labs, Berkeley, CA 94720}

\email{shiro@berkeley.edu}
 
\begin{abstract}

We report measurements of the cosmic microwave background (CMB) power spectrum from the complete 2008 South Pole Telescope (SPT) data set.
We analyze twice as much data as the first SPT power spectrum analysis, using an improved cosmological parameter estimator which fits multi-frequency models to the SPT $150$ and $220\,$GHz bandpowers.   
We find an excellent fit to the measured bandpowers with a model that includes lensed primary CMB anisotropy, secondary thermal (tSZ) and kinetic (kSZ) Sunyaev-Zel'dovich anisotropies, unclustered synchrotron point sources, and clustered dusty point sources.
In addition to measuring the power spectrum of dusty galaxies at high signal-to-noise, the data primarily constrain a linear combination of the kSZ and tSZ anisotropy contributions at $150\,$GHz and $\ell=3000$: $D^{tSZ}_{3000} + 0.5\,D^{kSZ}_{3000} = 4.5\pm 1.0 \,\mu{\rm K}^2$.
The $95\%$ confidence upper limits on secondary anisotropy power are $D^{tSZ}_{3000} < 5.3\,\mu{\rm K}^2$ and $D^{kSZ}_{3000} < 6.5\,\mu{\rm K}^2$.
We also consider the potential correlation of dusty and tSZ sources, and find it incapable of relaxing the tSZ upper limit.
These results increase the significance of the lower than expected
tSZ amplitude previously determined from SPT power spectrum measurements.
We find that models including non-thermal pressure support in groups and clusters predict tSZ power in better agreement with the SPT data.
Combining the tSZ power measurement with primary CMB data halves the statistical uncertainty on $\sigma_8$.
However, the preferred value of $\sigma_8$ varies significantly between tSZ models. 
Improved constraints on cosmological parameters from tSZ power
spectrum measurements require continued progress in the modeling of the tSZ power.  
\end{abstract}


\bigskip\bigskip

\section{Introduction}
\label{sec:intro}

Measurements of temperature anisotropy in the cosmic microwave background (CMB) have proven to be some of the most powerful and robust tests of cosmological theory \citep{komatsu10,larson10, reichardt09a, dunkley10}. 
The large- and intermediate-scale anisotropy is dominated by the
signal from the primary anisotropy of the CMB.  
On smaller
scales, the primary anisotropy is exponentially suppressed by Silk
damping \citep{silk68}; fluctuations on these scales are dominated by
foregrounds and secondary anisotropies.
These secondary
anisotropies are produced by interactions between CMB photons 
and matter that lies between the surface of last scattering and the
observer.  
Only recently have experiments reached sufficient
sensitivity and resolution to explore the cosmological information
encoded in the secondary CMB anisotropies.

The largest of these secondary anisotropies is expected to be the Sunyaev-Zel'dovich (SZ) effect \citep{sunyaev72}, which consists of two components: the thermal SZ (tSZ) and kinetic SZ (kSZ) effects. 
The tSZ effect occurs when CMB photons are inverse Compton scattered by hot electrons in the gravitational potential well of massive galaxy clusters. 
The resulting spectral distortion of the CMB blackbody spectrum creates a decrement in intensity at low frequencies, an increment at high frequencies, and a null near 220$\,$GHz.
At 150$\,$GHz, the
anisotropy induced by the tSZ effect is expected to dominate over the
primary CMB fluctuations at multipoles
$\ell\gtrsim3500$. 

The anisotropy power due to the tSZ effect depends sensitively on the normalization
of the matter power spectrum, as parametrized by the RMS of the mass distribution on 8h$^{-1}$ Mpc
scales, $\sigma_8$.
Measurements of the tSZ power spectrum have the potential to 
provide independent constraints on cosmological parameters such as $\sigma_8$
and improve the precision with which they are determined.
In principle, the tSZ power spectrum from large surveys can also be used to constrain non-standard cosmological models, for example, placing limits on the range of allowed early dark energy models  \citep{alam10}.

The tSZ power spectrum is challenging to model accurately because it includes significant contributions from galaxy clusters that span a wide range of mass and redshift. 
Modeling uncertainties arise from both non-gravitational heating effects in low-mass clusters and the limited body of observational data
on high-redshift clusters.
As a result, it has been difficult to accurately model the shape and
amplitude of the tSZ power spectrum expected for a given cosmology.
Recent models that vary in their treatment of cluster gas physics differ in amplitude by up to 50\% \citep{shaw10, trac10}.

The kSZ effect occurs when CMB photons are Doppler shifted by the bulk velocity of electrons in intervening gas.
The kSZ power spectrum is not weighted by the gas temperature and is, therefore, less sensitive to the non-linear effects that complicate tSZ models.
It does, however, depend on the details of reionization, which are not yet well understood.
In the standard picture of reionization, ionized bubbles form around the first stars or quasars, eventually merging and leading to a fully ionized Universe. 
The ionized bubbles will impart a Doppler shift on scattered CMB photons proportional to the bubble velocity.
This so-called ``patchy'' reionization kSZ signal depends on the duration of reionization as well as the bubble sizes \citep{gruzinov98,knox98}.
A detection of the kSZ power spectrum or upper limit on its amplitude can, in principle, lead to interesting constraints on the epoch 
of reionization \citep{zahn05}.

The first SPT power spectrum results were recently reported by \citet[hereafter L10]{lueker10}.  
This power spectrum was produced 
with 150 and $220\,$GHz data taken by the South Pole Telescope (SPT) on
$100\,{\rm deg}^2$  of sky.
The two frequencies were combined to remove foregrounds with a dust-like spectrum, resulting in the detection of
a linear combination of kSZ and tSZ power.
L10 reported the best-fit amplitude of the SZ power ($D_{\ell}=C_{\ell}\,\ell(\ell+1)/2\pi$) at a multipole of $\ell=3000$ to be 
$D^{tSZ}_{3000} + 0.46D^{kSZ}_{3000} = 4.2\pm1.5\,\mu{\rm K}^2$.
This amplitude was less than expected for many models, implying 
that either $\sigma_8$ was at the low end of the range allowed by other measurements, or the models were over-predicting
the tSZ power. 
The SPT $150\,$GHz bandpowers were used by \citet[hereafter H10]{hall10} to set a 95\% CL upper limit of $13\,\mu{\rm K}^2$ on the kSZ power at $\ell=3000$.

The Atacama Cosmology Telescope (ACT) collaboration has also measured the high-$\ell$ power spectrum \citep{das10}.
\citet{dunkley10} use the \citet{das10} bandpowers  to measure  the sum of the tSZ and kSZ power at $\ell=3000$ and $148\,$GHz to be $6.8 \pm 2.9 \,\mu {\rm K}^2$. 
The SPT and ACT bandpowers and resulting constraints on SZ power are consistent within the reported
uncertainties.

Since the publication of L10, 
several new models for the tSZ power spectrum have been published \citep{trac10,shaw10,battaglia10}. 
Most feature similar angular scale dependencies (indistinguishable with current data); however, the predicted amplitude of the SZ signal varies considerably between models.
The amplitude is generally reduced as the amount of star-formation, feedback, and non-thermal pressure support in clusters increases.
These effects can reduce the predicted tSZ power by 50\% compared to  the \citet[S10]{sehgal10} model used by L10, and greatly reduce the tension between the measured and predicted tSZ power.
 
In addition to the tSZ and kSZ signal, the measured power at these
frequencies and angular scales includes contributions from several
significant foregrounds. 
The power from dusty star-forming galaxies (DSFGs) has both a Poisson and clustered component, with distinct angular scale dependencies.  
Measurements of DSFG power at multiple millimeter wavelengths can be combined to constrain the DSFG spectral index.
H10 constrained the spectral index of the Poisson term to be $\alpha_p = 3.86 \pm 0.23$ and the index of the clustered term to be $\alpha_c = 3.8 \pm 1.3$ using fits to the L10 single-frequency bandpowers. 
H10 also interpret the implications of the spectral index constraint on the dust temperature and redshift distribution of the dusty galaxies.
\citet{dunkley10} also detect significant power attributed to clustered DSFGs in the ACT data. 
They find a preferred spectral index of $3.69 \pm 0.14$, consistent with the H10 result.

In this work, we make two key improvements upon the first SPT power spectrum release (L10 and H10). 
First, we include both fields observed by SPT in 2008, doubling the sky coverage and reducing the bandpower uncertainties by 30\%.
Second, we test cosmological models with a true multi-frequency analysis of the bandpowers,
properly accounting for the multi-frequency covariance matrix and frequency dependence of each component to estimate cosmological parameters.
Including the additional frequency information in the parameter estimation leads to improved model constraints.
We present the minimal model required to explain the data, and then explore  a number of extensions to this minimal model such as the possibility of the tSZ signal being correlated with dust emission.
None of these extensions significantly improve the fits or change the general conclusions concerning the amplitudes
of the kSZ and tSZ power.

We describe the instrument, observations, beams, and calibration strategy in \S\ref{sec:obs-reduc}. 
The time-ordered data  filtering and map-making algorithm is outlined in \S\ref{sec:analysis}, along with the procedure to derive bandpowers from maps. 
The results of tests for systematic errors applied to the SPT data are discussed in \S\ref{sec:jackknives}. 
The bandpowers are given in \S\ref{sec:results}, and the model is presented in \S\ref{sec:model}. 
The fitted parameters and their cosmological interpretation are given in \S\ref{sec:params} and \S\ref{sec:sigma8}, respectively.
We summarize our conclusions in \S\ref{sec:conclusions}.

\section{Instrument and Observations}
\label{sec:obs-reduc}
The SPT is an off-axis Gregorian telescope with a 10 meter diameter
primary mirror located at the South Pole.  
The receiver is equipped with 960 horn-coupled spiderweb bolometers with 
superconducting transition edge sensors.
In 2008, the focal plane included detectors at two frequency bands centered at
approximately $150$ and $220\,$GHz.  The telescope and receiver are discussed in more detail in \citet{ruhl04}, \citet{padin08}, and \citet{carlstrom09}. 

In this work, we use data at 150 and 220$\,$GHz from two $\sim$100 deg$^2$ fields
observed by SPT in the 2008 austral winter. 
The fields are centered at right ascension (RA) $5^\mathrm{h} 30^\mathrm{m}$, 
declination (decl.) $-55^\circ$ (J2000) (henceforth the \five\ field) and RA $23^\mathrm{h} 30^\mathrm{m}$, decl. $-55^\circ$ (the \twentythree\ field).
The locations of the fields were chosen for 
overlap with the optical Blanco Cosmology Survey (BCS)\footnote{http://cosmology.illinois.edu/BCS} and low dust emission observed by IRAS at 100$\,\mu$m \citep{schlegel98}.  
After data quality cuts, a total of 1100 hours of observations are used in this analysis.
 The
final map noise is 18$\,\mu\textrm{K}$-arcmin\footnote{Throughout this work, the unit $\textrm{K}$ refers to equivalent fluctuations in the CMB temperature, i.e.,~the temperature fluctuation of a 2.73$\,$K blackbody that would be required to produce the same power fluctuation.  
The conversion factor is given by the derivative of the blackbody spectrum, $\frac{dB}{dT}$, evaluated at 2.73$\,$K.} 
at 150$\,$GHz and 40$\,\mu\textrm{K}$-arcmin at 220$\,$GHz. 
These data include the majority of the sky area observed in 2008. 
In 2009 and 2010, the SPT has been used to survey an additional 1300 deg$^2$ to the same depth at $150\,$GHz.

The fields were observed with two different scan strategies.
The \five\ scan strategy involves
constant-elevation scans across the 10$^\circ$ wide field. 
After each scan back
and forth in azimuth across the field, the telescope executes a 0.125$^\circ$ step
in elevation.
A complete set of scans covering the entire field takes approximately two hours, and we
refer to each complete set as an observation. 

The \twentythree\ field was observed using a lead-trail scan strategy.
Two consecutive observations each map approximately half the field.
Due to the Earth's rotation, both halves of the field are observed at the same range of azimuth angle.
This would allow the removal of ground-synchronous signals if detected; however, we see no evidence for such a signal at the angular scales of interest.
 A complete observation of one half-field takes approximately 40 minutes.
The two halves are generally observed directly after one another, though for this analysis we also include a small number of lead trail pairs (31 out of 480)  which were not observed sequentially, due to poor weather or other interruptions.  
The requirement that both halves of the \twentythree\ field are observed at the same azimuth leads to a larger elevation step (0.268$^\circ$) and therefore less uniform coverage across the map.
We ameliorate this non-uniform coverage in the cross-spectrum analysis by combining two lead-trail map pairs into one `observation' unit.
The lead-trail pairs are chosen to to maximize coverage uniformity in the resulting map and minimize temporal offsets.

\subsection{Beam Functions}
\label{sec:beam}

\begin{figure}[t]\centering
\includegraphics[width=0.45\textwidth]{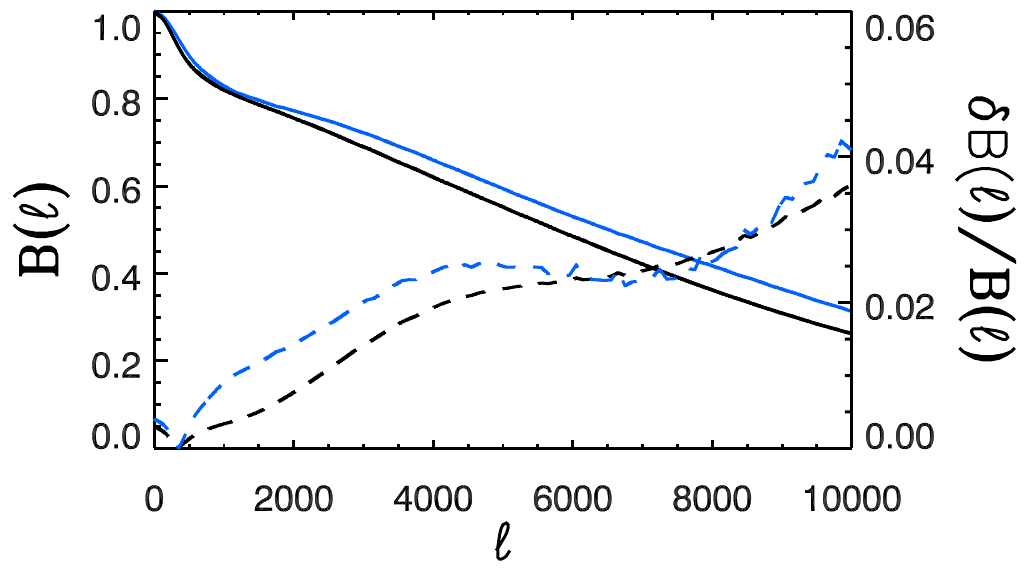}
  \caption[]{ {\it Left axis:} The measured SPT beam functions at $150\,$GHz ({\bf black line}) and $220\,$GHz ({\bf blue line}). {\it Right axis:} The fractional beam uncertainties at $150\,$GHz ({\bf black dashed line}) and $220\,$GHz ({\bf blue dashed line}). The beam uncertainty is parametrized by a three component model, the quadrature sum of which is plotted here.}
  \label{fig:beam}
\end{figure}

The power spectrum analysis presented here depends on an accurate measurement of the beam function, which is the azimuthally averaged Fourier transform of the beam map. 
Due to the limited dynamic range of the detectors, the SPT beams for the 2008 observing season were measured by combining maps of three sources:
Jupiter, Venus, and the brightest point source in each field.
Maps of Venus are used to stitch together the outer and inner beam maps from Jupiter and the point source, respectively.
Maps of Jupiter at radii $> 4^\prime$ are used to constrain a diffuse, low-level sidelobe that accounts for roughly 15\% of the total beam solid angle.
The beam within a radius of $4^\prime$ is measured on the brightest point source in the field. 
The effective beam is slightly enlarged by the effect of random errors in the pointing reconstruction.
We include this effect by measuring the main beam from a source in the final map. 
The main-lobe beam is approximately fit by 2D Gaussians with full width at half-maximum 
(FWHM) equal to $1.15^\prime$ at $150\,$GHz and $1.05^\prime$ at $220\,$GHz.
We have verified that the measured beam is independent of the point source used, although the signal-to-noise drops for the other (dimmer) sources in either field.
The beam measurement is similar to that described in L10, although the radii over which each source contributes to the beam map have changed.
However, the estimation of beam uncertainties has changed.
We now determine the top three eigenvectors of the covariance matrix for the beam function at each frequency, and include these as beam errors in the parameter fitting. 
These three modes account for $>99$\% of the beam covariance power.
The beam functions and the quadrature sum of the three beam uncertainty parameters are shown in Figure \ref{fig:beam}.

\subsection{Calibration}
\label{sec:calibration}

The absolute calibration of the SPT data is based on a comparison of the CMB power at degree scales in the WMAP 5-year maps with dedicated SPT calibration maps. 
These calibration maps cover four large fields totaling $1250\,{\rm deg}^2$ which were observed to shallow depth during the 2008 season.
Details of the cross-calibration with WMAP are given in L10, and remain essentially unchanged in this work.
Although the beam functions changed slightly at $\ell > 1000$ as a result of the improved beam measurement described in \S\ref{sec:beam}, this change had a negligible effect at the angular scales used in the WMAP calibration.
The calibration factors are therefore unchanged from L10; however as discussed above, we have slightly changed the treatment of beam and calibration errors, and are no longer folding part of the beam uncertainty into a calibration uncertainty.

We estimate the uncertainty of the temperature calibration factor to be $3.5\%,$ which is slightly smaller than the value of $3.6\%$ presented in L10.
We applied this calibration to the $220\,$GHz band by comparing $150\,$GHz 
and $220\,$GHz estimates of CMB anisotropy in the survey regions.  This internal 
cross-calibration for $220\,$GHz is more precise than, but consistent with, a direct absolute calibration using observations of RCW38.
We estimate the $220\,$GHz temperature calibration uncertainty to be $7.1\%$.  Because the $220\,$GHz 
calibration is derived from the $150\,$GHz calibration to WMAP, the calibration uncertainties 
in the two bands are correlated with a correlation coefficient of approximately 0.5.

As discussed in \S\ref{sec:baseline}, our MCMC chains also include two parameters to adjust the overall temperature calibration in each SPT band.
The resulting calibration factors are listed in the Table~\ref{tab:bandpowerssinglefreq} comments.

\section{Analysis}\label{sec:analysis}

In this section, we present an overview of the pipeline used to process the time-ordered data (TOD) to bandpowers. 
The method closely follows the approach used by L10, and we refer the reader to L10 for complete details of the method.
We highlight any differences between this analysis and that work.

Given the small field sizes, we use the flat-sky approximation. 
We generate maps using the oblique Lambert equal-area azimuthal projection and analyze these maps using Fourier transforms. 
Thus, the discussion of filtering and data-processing techniques refer to particular modes by their corresponding angular wavenumber {\bf k} in radians with $| {\bf k}| = \ell$.

\subsection{TODs to Maps}
\label{sec:mapping}

Each detector in the focal plane measures the CMB brightness
temperature plus noise, and records this measurement as the TOD.
Noise in the TOD contains contributions from the instrument and atmosphere. 
The instrumental noise is largely uncorrelated between detectors. 
Conversely, the atmospheric contribution is highly
correlated across the focal plane, because the beams of individual detectors overlap significantly as they pass through turbulent layers in the atmosphere.

We bandpass filter the TOD to remove noise outside the signal band.
The TOD are low-pass filtered at 12.5 Hz ($\ell \sim 18000$) to remove noise above the Nyquist frequency of the map pixelation. 
The TOD are effectively high-pass filtered by the removal of a Legendre polynomial from the TOD of each detector on a by-scan basis. 
The order of the polynomial is chosen to have the same number of degrees of freedom (dof) per unit angular distance ($\sim$1.7 dof per degree).
The polynomial fit removes low frequency noise contributions from the instrument and atmosphere.

Correlated atmospheric signals remain in the TOD after the bandpass filtering.
We remove the correlated signal by subtracting the mean signal over each bolometer 
wedge\footnote{The SPT array consists of 6 pie shaped bolometer 
wedges of 160 detectors, of which five wedges are used in this analysis. 
Wedges are configured with a set of filters that determine their observing frequency (e.g. 150 or $220\,$GHz).}
 at each time sample. 
This filtering scheme is slightly different from that used by L10, although nearly identical in effect at $150\,$GHz. 
L10 subtracted a plane from all wedges at a given observing frequency at each time sample. Since SPT had two wedges of $220\,$GHz bolometers and three wedges of $150\,$GHz bolometers in 2008, the L10 scheme filters the $220\,$GHz data more strongly than the $150\,$GHz data. The wedge-based common mode removal implemented in this work results in more consistent filtering between the two observing bands. 

The data from each bolometer is inverse noise weighted based on the calibrated, pre-filtering detector power spectral density.
We bin the data into map pixels based on pointing information, and in the case of the \twentythree\ field, combine two pairs of lead and trail maps to form individual observations as discussed in \S\ref{sec:obs-reduc}.
A total of 300 and 240 individual observation units are used in the subsequent analysis of the \five\ and \twentythree\ fields, respectively.

\subsection{Maps to Bandpowers}
\label{sec:bandest}

We use a pseudo-$C_\ell$ method to estimate the bandpowers.
 In pseudo-$C_\ell$ methods, bandpowers are estimated directly from the Fourier transform of the map after correcting for effects such as TOD filtering, beams, and finite sky coverage. 
We process the data using a cross spectrum based analysis \citep{polenta05, tristram05} in order to eliminate noise bias.  
Beam and filtering effects are corrected for according to the formalism in the MASTER algorithm \citep{hivon02}.  
We report the bandpowers in terms of $\mathcal{D}_\ell$, where
\begin{equation}
\mathcal{D}_\ell=\frac{\ell\left(\ell+1\right)}{2\pi} C_\ell\;.
\end{equation}

The first step in the analysis is to calculate the Fourier transform $\tilde{m}^{(\nu_i,A)}$ of the map for each frequency $\nu_i$ and observation $A$. 
All maps of the same field are apodized by a single window that is chosen to mask out detected point sources and avoid sharp edges at the map borders.
After windowing, the effective sky area used in this analysis is $210\,$deg$^2$.
Cross-spectra are then calculated for each map-pair on the same field;
a total of 44850 and 28680 pairs are used in the \five and \twentythree fields, respectively.
We take a weighted average of the cross-spectra within an $\ell$-bin, $b$,
\begin{equation}
\label{eqn:ddef}
 D^{\nu_i\times\nu_j, AB}_b\equiv \left< \frac{{\rm k}({\rm k}+1)}{2\pi}\tilde{m}^{(\nu_i,A)}_\textbf{k} \tilde{m}^{(\nu_j,B)*}_\textbf{k} \right>_{k \in b}.
\end{equation}
As in L10, we find a simple, uniform selection of modes at $k_x > 1200$ is close to the optimal mode-weighting.
With the adopted flat-sky approximation, ${\bf \ell} = k$.
 
 The bandpowers above are band-averaged pseudo-$C_\ell$s that can be related to the true astronomical power spectrum $\widehat{D}$ by
 \begin{equation}
\widehat{D}^{\nu_i\times\nu_j,AB}_b\equiv \left(K^{-1}\right)_{bb^\prime}D^{\nu_i\times\nu_j,AB}_{b^\prime}.
\end{equation}
The $K$ matrix accounts for the effects of timestream filtering, windowing, and band-averaging. 
This matrix can be expanded as
\begin{equation}
\label{eqn:kdef}
K^{\nu_i\times\nu_j}_{bb^\prime}=P_{bk}\left(M_{kk^\prime}[\textbf{W}]\,F^{\nu_i\times\nu_j}_{k^\prime}B^{\nu_i}_{k^\prime}B^{\nu_j}_{k^\prime}\right)Q_{k^\prime b^\prime}.
\end{equation}
Here $Q_{kb}$ and $P_{bk}$ are the binning and re-binning operators \citep{hivon02}. $B^{\nu_i}_{k}$ is the beam function for frequency $\nu_i$, and $F^{\nu_i\times\nu_j}_{k}$ is the $k$-dependent transfer function which accounts for the filtering and map-making procedure. The mode-mixing due to observing a finite portion of the sky is represented by $M_{kk^\prime}[\textbf{W}]$, which is calculated analytically from the known window $\textbf{W}$.

\subsubsection{Transfer Function}

We calculate a transfer function for each field and observing frequency.
The transfer functions are calculated from simulated observations of 300 sky realizations that have been smoothed by the appropriate beam. 
Each sky realization is a Gaussian realization of the best-fit lensed WMAP7 $\Lambda$CDM model plus a Poisson point source contribution. 
The Poisson contribution is selected to be consistent with the results of H10, and includes radio source and DSFG populations.
 The amplitude of the radio source term is set by the \citet{dezotti05} model source counts at $150\,$GHz with an assumed spectral index of $\alpha_r=-0.5$. 
 The DSFG term has an assumed spectral index of 3.8.
 The effective point source powers are $C_\ell^{\rm 150x150} = 7.5\times10^{-6} \,\mu{\rm K}^2$, $C_\ell^{\rm 150x220} = 2.3\times10^{-5} \,\mu{\rm K}^2$, and $C_\ell^{\rm 220x220} = 7.8\times10^{-5} \,\mu{\rm K}^2$. 
The sky realizations are sampled using the SPT pointing information, filtered identically to the real data, and processed into maps.  
The power spectrum of the simulated maps is compared to the known input spectrum, $C^{\nu,\textrm{theory}}$, to calculate the effective transfer function \citep{hivon02} in an iterative scheme,
\begin{equation}\label{eqn:transiter}
F^{\nu, (i+1)}_{k}=F^{\nu,(i)}_{k}+\frac{\left<D^{\nu}_{k}\right>_{\rm MC} -  M_{kk^\prime} F_{k^\prime}^{\nu,(i)} {B^\nu_{k^\prime}}^2 C^{\nu,\textrm{theory}}_{k^\prime}}{{B^\nu_{k}}^2 C^{\nu,\textrm{theory}}_{k} w_2}.
\end{equation}
Here $w_2= \int d\!x \textbf{W}^2$ is a normalization factor for the area of the window. 
We find that the transfer function estimate converges after the first iteration and use the fifth iteration.
For both fields and both bands, the result is a slowly varying function of $\ell$ with values that range from a minimum of 0.6 at $\ell=2000$ to a broad peak near $\ell=5000$ of 0.8.

\subsubsection{Bandpower Covariance Matrix}
The bandpower covariance matrix includes two terms: variance of the signal in the field (cosmic variance), and instrumental noise variance.  
The signal-only Monte Carlo bandpowers are used to estimate the cosmic variance contribution.
The instrumental noise variance is calculated from the distribution of the cross-spectrum bandpowers $D^{\nu_i\times\nu_j,AB}_b$ between observations A and B, as described in L10. 
We expect some statistical uncertainty of the form  
\begin{equation}
\left<\left(\textbf{c}_{ij}-\left<\textbf{c}_{ij}\right>\right)^2\right>=\frac{\textbf{c}_{ij}^2+\textbf{c}_{ii}\textbf{c}_{jj}}{n_{obs}}
\end{equation}
in the estimated bandpower covariance matrix.
Here, $n_{obs}$ is the number of observations that go into the covariance estimate.
This uncertainty on the covariance is significantly higher than the true covariance for almost all off-diagonal terms due to the dependence of the uncertainty on the (large) diagonal covariances.
We reduce the impact of this uncertainty by ``conditioning'' the covariance matrix. 

How we condition the covariance matrix is determined by the  form we expect it to assume.
The covariance matrix can be viewed as a set of nine square blocks, with the three on-diagonal blocks corresponding to the covariances of a $150\times150$, $150\times220$, or $220\times220\,$GHz spectrum.
Since the bandpowers reported in Tables \ref{tab:bandpowerssinglefreq} and \ref{tab:bandpowersdif} are obtained by first computing power spectra and covariance matrices for bins of width $\Delta\ell=100$ with a total of 80 initial $\ell$-bins, each of these blocks is an $80 \times 80$ matrix. 
The shape of the correlation matrix in each of these blocks is expected to be the same, as it is set by the apodization window.   
As a first step to conditioning the covariance matrix, we calculate the correlation matrices for the three on-diagonal blocks and average all off-diagonal elements at a fixed separation from the diagonal in each block,
\begin{equation}
\label{eqn:covcond}
\textbf{c}^\prime_{kk^\prime}=\frac{
\sum_{k_1-k_2=k-k^\prime} 
\frac{\widehat{\textbf{c}}_{k_1k_2}}{\sqrt{\widehat{\textbf{c}}_{k_1k_1}\widehat{\textbf{c}}_{k_2k_2}}}
}{\sum_{k_1-k_2=k-k^\prime} 1}.
\end{equation}
This averaged correlation matrix is then applied to all nine blocks. 

The covariance matrix generally includes an estimate of both signal and noise variance.
However, the covariance between the $150 \times 150$ and $220\times220$ bandpowers is a special case:
since we neither expect nor observe correlated noise between the two frequencies in the signal band,
we include only the signal variance in the two blocks describing their covariance.

\subsubsection{Combining the Fields}
We have two sets of bandpowers and covariances---one per field---which must be combined to find the best estimate of the true power spectrum.
We calculate near-optimal weightings for this combination at each frequency and $\ell$-bin using the diagonal elements of the covariance matrix, 
\begin{equation}
w^{\nu_i\times\nu_j}_b \propto 1/(C^{(\nu_i \times \nu_j) \times (\nu_i \times \nu_j)}_{bb}).
\end{equation}
This produces a diagonal weight matrix, $w$, for which the bandpowers can be calculated:
\begin{equation}
\label{eqn:combineband}
\widehat{D}^{\nu_i\times\nu_j}_b = \sum_{\zeta} \left(w_\zeta^{\nu_i\times\nu_j} \widehat{D}_\zeta^{\nu_i\times\nu_j}\right)_b,
\end{equation}
where $\zeta$ denotes the field.
The covariance will be
$$\textbf{c}^{(\nu_i \times \nu_j) \times (\nu_m \times \nu_n) }_{bb^\prime} =$$
\begin{equation}
\label{eqn:combinecov}
=\sum_{\zeta} \left(w_\zeta^{\nu_i\times\nu_j} \textbf{c}_\zeta^{(\nu_i \times \nu_j) \times (\nu_m \times \nu_n)} w_\zeta^{\nu_m\times\nu_n}\right)_{bb^\prime}. 
\end{equation}
After combining the two fields, we average the bandpowers and covariance matrix into the final $\ell$-bins. Each initial $\Delta\ell=100$ sub-bin gets equal weight.

\section{Jackknife tests}
\label{sec:jackknives}

We apply a set of jackknife tests to the data to search for
possible systematic errors. 
In a jackknife test, the data set is divided into two halves
based on features of the data associated with potential sources of systematic error.
After differencing the two halves to remove any astronomical
signal, the resulting power spectrum is compared to
zero. Significant deviations from zero would 
indicate a systematic problem or possibly a noise misestimate.  
Jackknives with a cross-spectrum (as opposed to an auto-spectrum) estimator are less sensitive to small noise misestimates since there is no noise bias term to be subtracted.
We implement the jackknives in the cross-spectrum framework by differencing single pairs of observations and applying the cross-spectrum estimator outlined in \S\ref{sec:bandest} to the set of differenced pairs.   
In total, we perform three jackknife tests based on the observing parameters:  time, scan direction, and azimuthal range.

\begin{figure*}[ht]\centering
\includegraphics[width=0.8\textwidth]{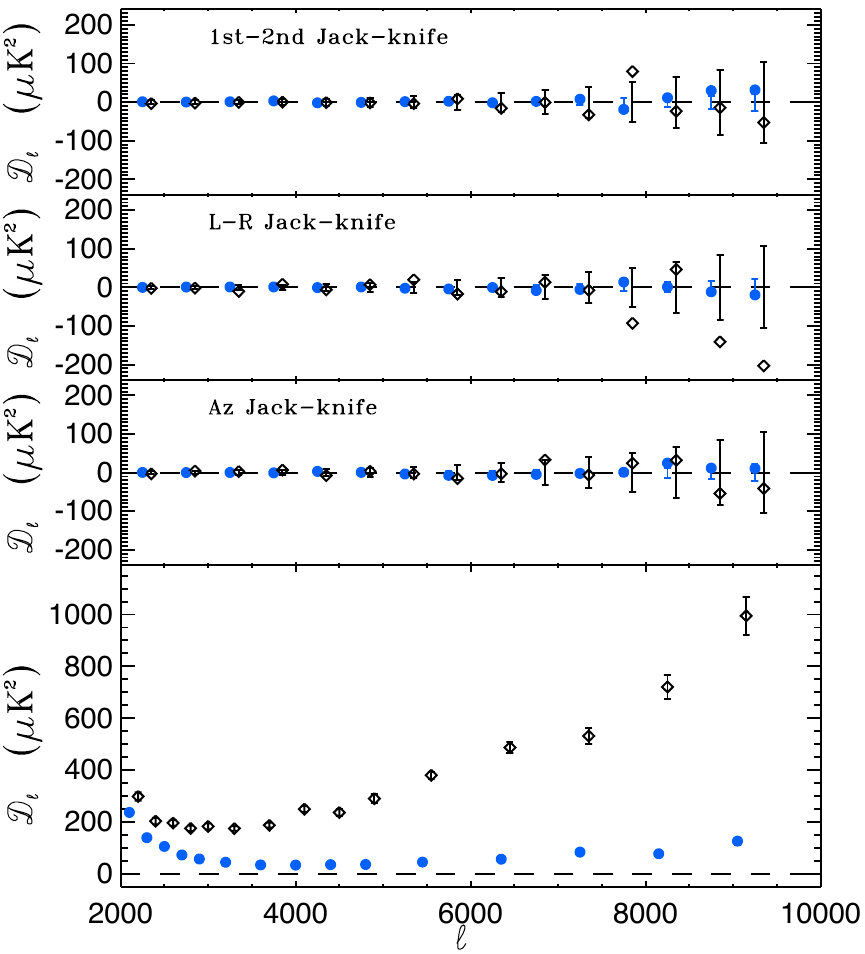}
  \caption[]{ Jackknives for the SPT data set at $150\,$GHz ({\bf blue circles}) and $220\,$GHz ({\bf black diamonds}). 
  For clarity, the $220\,$GHz jackknives have been shifted to the right by $\Delta\ell = 100$. 
  {\it Top panel:} Bandpowers of the ``first half - second half'' jackknife compared to the expected error bars about zero signal. 
  Disagreement with zero would indicate either a noise misestimate or a time-dependent systematic signal.   
  {\it Second panel:} Power spectrum of the left-going minus right-going difference map. 
  This test yields strong constraints on the accuracy of the detector transfer function deconvolution and on possible directional systematics. 
  {\it Third panel:} Bandpowers for the difference map when the data are split based on the observed very large scale ($\ell < 100$) ground pickup as a function of azimuth. 
  Signals fixed in azimuth, such as ground pickup on smaller scales, would produce non-zero power.  
  We see no evidence for ground-based pickup across this $\ell$-range. The cumulative probability to exceed the $\chi^2$ observed in these three tests at 150 and $220\,$GHz is 55\%. {\it Bottom panel:} The un-differenced SPT power spectra at each frequency for comparison.}
\label{fig:jack}
\end{figure*}

The data can be split based on the time of observation to search for
variability in the calibration, beams, detector time constants,
or any other potentially time variable aspect of the observations. 
The ``first half - second
half'' jackknife probes variations on month time scales. 
Results for the ``first half - second
half'' jackknife are shown in the top panel of Figure \ref{fig:jack}. 
We note that in the \twentythree\ field, the combination of individual field observations into four-map units, discussed in section \ref{sec:obs-reduc}, 
results in a small number of constituent maps (9 out of 960) that are grouped into the wrong side of this split.
This does not affect the \five\ field, where the observation unit is a single map.

The data can also
be split based on the direction of the scan in a ``left\,-\,right''
jackknife (panel 2 of Figure \ref{fig:jack}). We would expect to see
residual power here if the detector transfer
function has been improperly de-convolved, if the telescope
acceleration at turn-arounds induces a signal through sky modulation or microphonics, or if
the wind direction is important.  

Sidelobe pickup could potentially introduce spurious
signals into the map from features on the ground. 
In deep coadds of unfiltered data, we observe features with scales of several degrees that are fixed 
with azimuth and are presumably caused by ground pickup.
This ground pickup is significantly reduced by the wedge common-mode removal, but could still
exist at low levels. 
We split the data as a function of azimuth based on the observed RMS signal on these large scales.
This split is different than the choice made in L10, which was based
 upon the distance in azimuth between an observation and the closest building.
The azimuthal jackknife is shown in the third panel of 
Figure \ref{fig:jack}.   

We calculate the $\chi^2$ with respect to zero for
each jackknife over the range $\ell \in [2000,9500]$ in bins with $\Delta\ell = 500$.
The probability to exceed the measured $\chi^2$ (PTE) for the
three individual jackknives is 5\%, 66\%, and 15\% at $150\,$GHz and 99\%, 21\%, and 98\% at $220\,$ GHz for the ``first half - second half,'' ``left\,-\,right,'' and ``azimuth'' jackknives, respectively.
The combined PTE for the three jackknives is 10\% for the $150\,$GHz data, 96\% 
for the $220\,$GHz data and 55\% for the combined set of both
frequencies.  We thus find no evidence for systematic contamination of the
SPT band powers. 

\section{Power spectra}
\label{sec:results}

\begin{figure*}[ht]\centering
\includegraphics[width=0.9\textwidth]{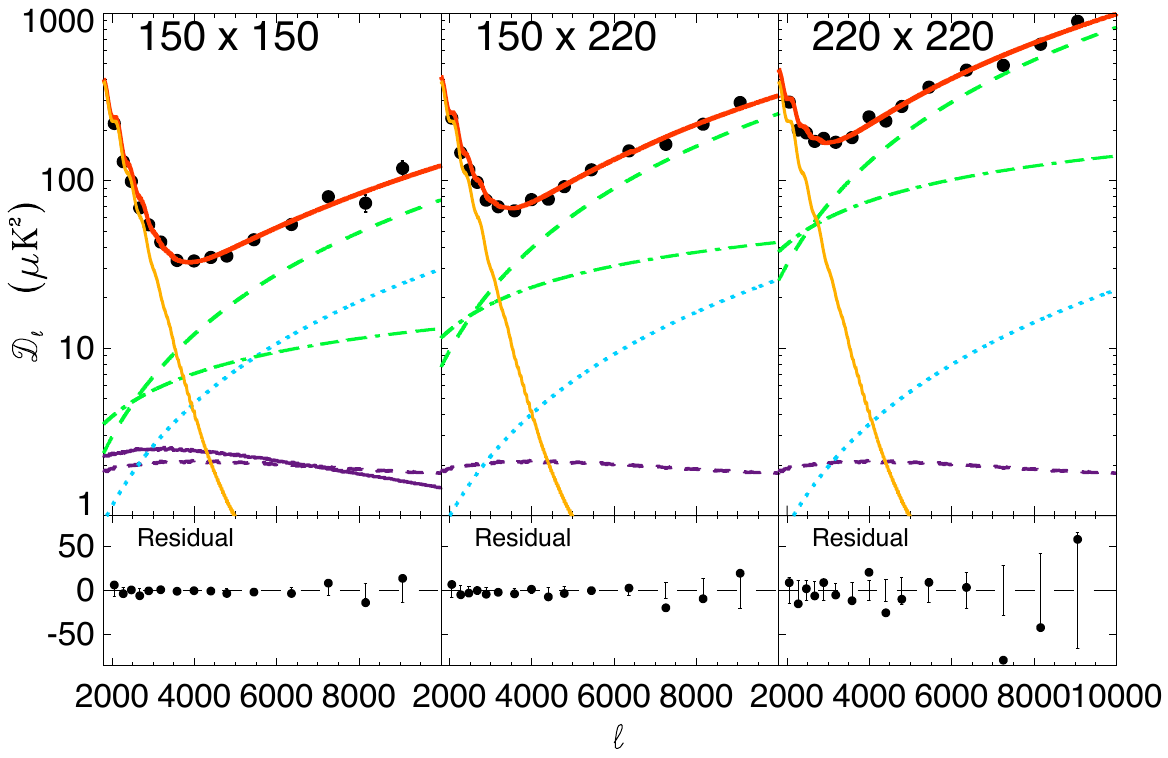}
  \caption[]{{\it Top panel:} From left to right, the SPT $150\,$GHz, $150\times220\,$GHz, and $220\,$GHz bandpowers.
  Overplotted is the best-fit model ({\bf red line}) with components shown individually.
  The lensed primary CMB anisotropy is marked by an {\bf orange line}. The best-fit tSZ ({\bf purple line})  and predicted kSZ ({\bf purple dashed line})  power spectra are also shown.
  The predicted radio source term is represented by the {\bf blue dots}.
  The DSFG Poisson term at each frequency is denoted by the {\bf green dashed line} and the clustered DSFG component by the {\bf green dot-dash line}.
  The damping tail of the primary CMB anisotropy is apparent below $\ell = 3000$. 
  Above $\ell = 3000$, there is a clear excess with an angular scale dependence consistent with point sources.  
  These sources have low flux (sources detected at $>$$\,5\,\sigma$ at $150\,$GHz have been masked) and a rising frequency spectrum, consistent with expectations for DSFGs.  
  {\it Bottom panel:} Plot of the residual between the measured bandpowers and best-fit spectrum.
}
\label{fig:clspt}
\end{figure*}

The bandpowers presented in Table \ref{tab:bandpowerssinglefreq} and plotted in Figure \ref{fig:clspt} are the result of applying the analysis described in \S\ref{sec:analysis} to the 210 deg$^2$ observed by SPT in 2008.
The errors shown in Table \ref{tab:bandpowerssinglefreq} are the diagonal elements of the covariance matrix. The full covariance matrix including off-diagonal elements can be found at the SPT website\footnote{http://pole.uchicago.edu/public/data/shirokoff10/}
along with the window functions describing the transformation from a theoretical spectrum to these bandpowers \citep{knox99}.

In these power spectra, we have measured the mm-wave anisotropy power
at $\ell > 2000$ with the highest significance to date.
Anisotropy power is detected at $92\,\sigma$ at $150\,$GHz and $74\,\sigma$ at $220\,$GHz. 
Below $\ell \sim 3000$, the bandpowers reflect the Silk damping tail of the primary CMB anisotropy.
At $\ell > 3000$, the bandpowers follow the ${\it D}_\ell \propto \ell^2$ form expected for a Poisson point source distribution.
The measured spectral index ($\alpha \sim 3.6$ where $S_\nu \propto \nu^\alpha$) points to a dust-like spectrum for the point sources indicating that the DSFGs are the dominant point source population.
However, the CMB and Poisson terms do not fully explain the data.
Adding two free parameters representing the tSZ and effects of point source clustering improves the best-fit model likelihood, $\mathcal{L}$, by $\Delta {\rm ln} \mathcal{L} = 65$.
Using a combination of the $150\,$GHz auto-spectrum, $150\times220$ cross-spectrum and $220\,$GHz auto-spectrum, we can separate these foregrounds and secondary anisotropies; 
however,  substantial degeneracies remain.
Future SPT data analyses will include a third frequency band, which we expect will significantly reduce these degeneracies.

\begin{table*}[ht!]
\begin{center}
\caption{\label{tab:bandpowerssinglefreq} Single-frequency bandpowers}
\small
\begin{tabular}{cc|cc|cc|cc}
\hline\hline
\rule[-2mm]{0mm}{6mm}
& &\multicolumn{2}{c}{$150\,$GHz} & \multicolumn{2}{c}{$150\times220\,$GHz} & \multicolumn{2}{c}{$220\,$GHz} \\
$\ell$ range&$\ell_{\rm eff}$ &$\hat{D}$ ($\mu{\rm K}^2$)& $\sigma$ ($\mu{\rm K}^2$) &$\hat{D}$ ($\mu{\rm K}^2$)& $\sigma$ ($\mu{\rm K}^2$)&$\hat{D}$ ($\mu{\rm K}^2$)& $\sigma$ ($\mu{\rm K}^2$)\\
\hline
2001 - 2200 & 2056 & 242.1 &   6.7 & 248.8 &   8.3 &  295.9 &  14.9 \\ 
2201 - 2400 & 2273 & 143.2 &   4.2 & 154.7 &   5.6 &  201.5 &  11.7 \\ 
2401 - 2600 & 2471 & 109.3 &   3.2 & 122.1 &   4.5 &  193.5 &  11.0 \\ 
2601 - 2800 & 2673 &  75.9 &   2.6 & 102.8 &   4.1 &  172.1 &  10.4 \\ 
2801 - 3000 & 2892 &  60.2 &   2.3 &  80.4 &   3.7 &  179.3 &  11.1 \\ 
3001 - 3400 & 3184 &  47.5 &   1.5 &  73.5 &   2.5 &  169.7 &   8.2 \\ 
3401 - 3800 & 3580 &  36.9 &   1.6 &  69.5 &   2.7 &  180.7 &   9.2 \\ 
3801 - 4200 & 3993 &  36.7 &   1.8 &  81.0 &   3.3 &  240.1 &  11.6 \\ 
4201 - 4600 & 4401 &  38.5 &   2.2 &  81.5 &   3.9 &  226.6 &  12.6 \\ 
4601 - 5000 & 4789 &  39.3 &   2.7 &  96.9 &   4.7 &  276.9 &  15.2 \\ 
5001 - 5900 & 5448 &  49.2 &   2.5 & 122.3 &   4.2 &  361.3 &  13.3 \\ 
5901 - 6800 & 6359 &  60.8 &   3.9 & 158.7 &   6.2 &  457.4 &  20.0 \\ 
6801 - 7700 & 7256 &  89.1 &   6.0 & 173.9 &   9.5 &  488.8 &  28.4 \\ 
7701 - 8600 & 8159 &  81.7 &   9.5 & 229.2 &  14.2 &  653.0 &  42.3 \\ 
8601 - 9500 & 9061 & 131.9 &  14.4 & 309.1 &  21.9 &  895.2 &  66.0 \\ 
\hline
\end{tabular}
\tablecomments{ Band multipole range and weighted value $\ell_{\rm eff}$, bandpower $\hat{D}$, 
and uncertainty $\sigma$ for the $150\,$GHz auto-spectrum, cross-spectrum, and $220\,$GHz auto-spectrum of the SPT fields. 
The quoted uncertainties include instrumental noise and the Gaussian sample variance of the primary CMB and the point source foregrounds. 
The sample variance of the SZ effect, beam uncertainty, and calibration uncertainty are not included. 
To include the preferred calibration from the MCMC chains, these bandpowers should be multiplied by 0.92, 0.95, and 0.98 at $150\,$GHz, $150\times220\,$GHz, and $220\,$GHz respectively. (See \S\ref{sec:baseline}.)
Beam uncertainties are shown in Figure~\ref{fig:beam} and calibration uncertainties are quoted in \S\ref{sec:calibration}.
Point sources above $6.4\,$mJy at $150\,$GHz have been masked out in this analysis. 
This flux cut substantially reduces the contribution of radio sources to the bandpowers, although DSFGs below this threshold contribute significantly to the bandpowers.  }
\normalsize
\end{center}
\end{table*}

\begin{figure*}[ht]\centering
\includegraphics[width=0.9\textwidth]{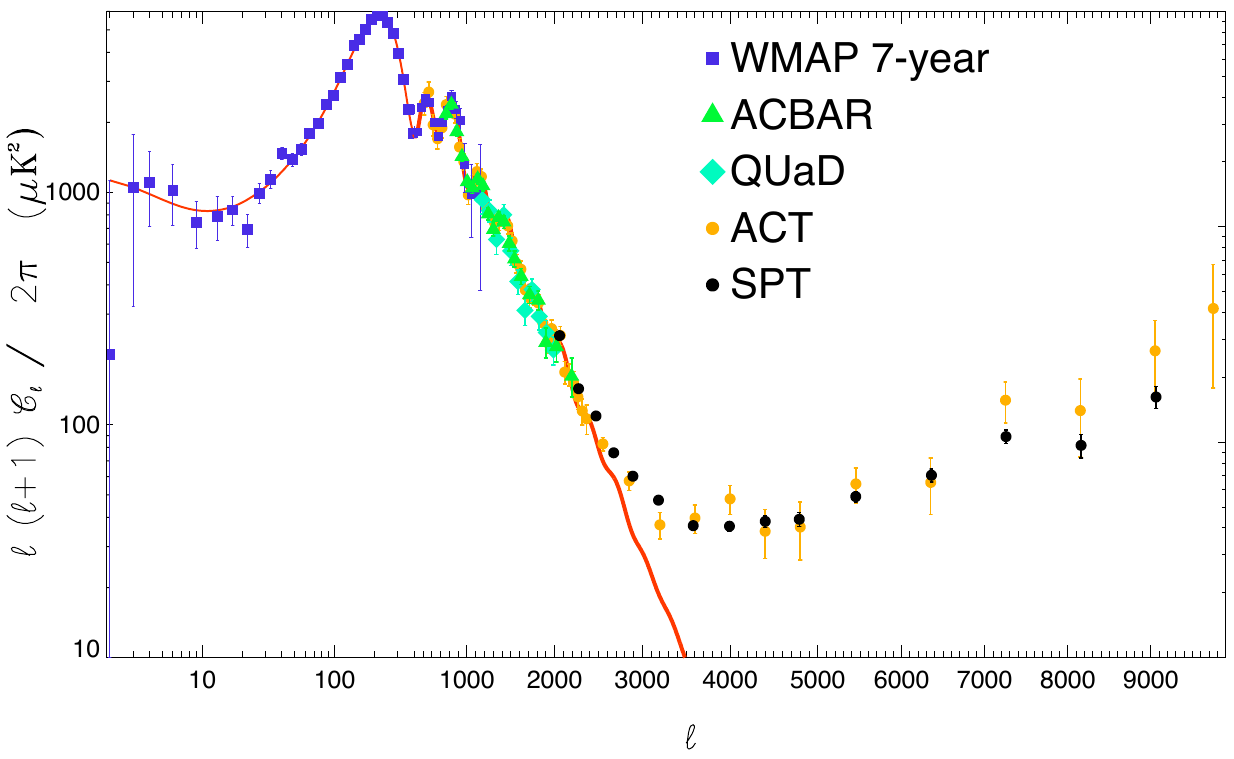}
  \caption[]{The SPT $150\,$GHz bandpowers ({\bf black circles}), WMAP7 bandpowers ({\bf purple squares}), ACBAR bandpowers ({\bf green triangles}), QUaD bandpowers ({\bf cyan diamonds}), and ACT $150\,$GHz bandpowers ({\bf orange circles}) plotted against the best-fit lensed $\Lambda$CDM CMB spectrum.
  The damping tail of the primary CMB anisotropy is apparent below $\ell = 3000$. 
  Above $\ell = 3000$, there is a clear excess due to secondary anisotropies and residual point sources that has now been measured by both SPT and ACT.
Note that the source masking threshold in the SPT data ($6.4\,$mJy) is lower than that in the ACT data, so we expect less radio source power at high $\ell$.
We have multiplied the SPT bandpowers by the best-fit calibration of 0.92 as determined in parameter fits.
  }
  \label{fig:clall}
\end{figure*}

\section{Cosmological model }
\label{sec:model}

We fit the SPT bandpowers to a model including lensed primary CMB anisotropy, secondary tSZ and kSZ anisotropies, galactic cirrus, an unclustered population of radio sources and a clustered population of DSFGs. 
In total, this model has six parameters describing the primary CMB anisotropy, up to two parameters describing secondary CMB anisotropies, and up to eleven parameters describing foregrounds.

Parameter constraints are calculated using the publicly available {\textsc CosmoMC}\footnote{http://cosmologist.info/cosmomc} package \citep{lewis02b}.
We have added two modules to handle the high-$\ell$ data; one to model the foregrounds and secondary anisotropies and one to calculate the SPT likelihood function.
These extensions, with some differences in the treatment of the secondary anisotropy, are also discussed in \citet{millea10}.
These modules and instructions for compiling them are available at the SPT website.\footnote{http://pole.uchicago.edu/public/data/shirokoff10/}

Although there are observational and theoretical reasons to expect
contributions to the power spectra from each of the components outlined
above, a good fit to the current SPT data can be obtained with a much
simpler, four-parameter extension to the primary CMB anisotropy.
This minimal model contains
four free parameters: the tSZ amplitude, the Poisson and clustered
DSFG amplitudes, and a single spectral index which sets the frequency
dependence of both DSFG terms.  
All other parameters are tightly constrained by external data.  
A similar model was used in recent ACT analysis \citep{dunkley10}.  
We adopt this four-parameter
extension as the baseline model for this work.

In this baseline model, several simplifications are made.  
The distributions of spectral indices are set to delta functions ($\sigma_X= 0$).  
The radio source spectral index is fixed to $\alpha_{r}=-0.53$.  
The amplitudes of the kSZ, radio source, and cirrus power spectra are fixed to expected values.  
The tSZ-DSFG correlation is set to zero.  
We discuss the priors placed on these components in the following subsections.

To compare the SPT data to these cosmological models, we combine the SPT high-$\ell$ bandpowers with low-$\ell$ CMB measurements from WMAP7, ACBAR, and QUaD \citep{larson10,reichardt09a,brown09}.
During the preparation of this paper, the ACT bandpowers were published \citep{das10}; however, it was not until after the 
model fitting runs were completed that the window functions were made available. 
For this reason, we do not include the ACT bandpowers in the fits although they are the best current constraints for $\ell \in [1600,2000]$. 
We expect that including the ACT data would have little impact on the results.
The low-$\ell$ CMB bandpowers tightly constrain the $\Lambda$CDM model space.
We restrict the ACBAR and QUaD bandpowers to $\ell < 2100$ where the details of the secondary anisotropies and point source spectra are negligible compared to the primary CMB anisotropy. 
These additional components are challenging to model between experiments since they are frequency-dependent and, in the case of point sources, mask-dependent.
In practice, removing this restriction does not affect the results since the SPT data dominates the constraints above $\ell = 2000$.

\subsection{Primary CMB Anisotropy} 
\label{sec:primarycmb}

We use the standard, six-parameter, spatially flat, lensed 
$\Lambda$CDM cosmological model to predict the primary CMB temperature anisotropy. 
The six parameters are the baryon 
density $\Omega_b h^2$, the density of cold dark matter $\Omega_c h^2$, the optical 
depth to recombination $\tau$, the angular scale of the peaks $\Theta$, the 
amplitude of the primordial density fluctuations $\ln[10^{10} A_s]$, and the 
scalar spectral index $n_s$.    

Gravitational lensing of CMB anisotropy 
by large-scale structure tends to increase the power at small angular scales, with the
potential to influence the high-$\ell$ model fits.  The
calculation of lensed CMB spectra out to $\ell=10000$ is
prohibitively expensive in computational time.  
Instead, we approximate the impact of lensing by adding a fixed lensing template calculated for the best-fit WMAP7 cosmology to unlensed CAMB\footnote{http://camb.info} spectra. 
The error due to this lensing approximation is less than $0.5\,\mu{\rm K}^2$ for the allowed parameter space at  $\ell > 3000$. 
This is a factor of 20 (260) times smaller than
the secondary and foreground power at 150 (220) GHz and should have an
insignificant impact on the likelihood calculation. 
We tested this assumption by including the full lensing treatment in a single chain with the baseline model described above.  
As anticipated, we found no change in the resulting parameter constraints. 

We find the data from SPT, WMAP7, ACBAR, and QUaD considered here prefer gravitational lensing at $\Delta {\rm ln} \mathcal{ L}=-5.9 ~(3.4\,\sigma)$. 
Most of this detection significance is the result of the low-$\ell$ data and not the high-$\ell$ SPT data presented here; without including SPT data in fits, lensing is preferred at $3.0\,\sigma$. 
In estimating the preference for gravitational lensing, we used the true lensing calculation instead of the lensing approximation described above.

\subsection{Secondary CMB Anisotropy}

We parametrize the secondary CMB anisotropy with two terms describing the amplitude of the tSZ and kSZ power spectra: $\amplitudeletter{tSZ}$ and $\amplitudeletter{kSZ}$.
The model for the SZ terms can be written as
\begin{equation}
D^{SZ}_{\ell,\nu_1,\nu_2} = \amplitudeletter{tSZ}  \frac{f_{\nu_1} f_{\nu_2} } { f_{\nu_0}^2} \frac{\modelletter^{tSZ}}{\modelnorm^{tSZ}} + \amplitudeletter{kSZ} \frac{ \modelletter^{kSZ}}{\modelnorm^{kSZ}} . 
\end{equation}
Here, $\modelletter^{X}$ denotes the theoretical model template for component X at frequency $\nu_0$. 
The frequency dependence of the tSZ effect is encoded in $f_\nu$; at the base frequency $\nu_0$, $D^{tSZ}_{3000,\nu_0,\nu_0} = \amplitudeletter{tSZ}$.
The kSZ effect has the same spectrum as the primary CMB anisotropy, so its amplitude is independent of frequency.
In this work, we set $\nu_0$ to be the effective frequency of the SPT $150$ GHz band (see \S\ref{subsec:efffreq}).
 
\subsubsection{tSZ Power Spectrum}
\label{sec:tszmodels}

We adopt four different models for the tSZ power
spectrum. Following L10, we use the power spectrum predicted by S10 as
the baseline model. S10 combined the semi-analytic model for the
intra-cluster medium (ICM) of \citet{bode09} with a cosmological N-body simulation
to produce simulated thermal and kinetic SZ maps from which the
template power spectra were measured. 
The assumed cosmological parameters are ($\Omega_b$, $\Omega_m$, $\Omega_\Lambda$, $h$, $n_s$, $\sigma_8$) = (0.044, 0.264, 0.736, 0.71, 0.96, 0.80).
At $\ell = 3000$, this model
predicts $D^{tSZ}_{3000}=7.5\, \mu{\rm K}^2$ of tSZ power in the SPT
$150\,$GHz band. We use this model in
all chains where another model is not explicitly specified.

We also consider tSZ power spectrum models reported by \citet{trac10},
\citet{battaglia10}, and \citet{shaw10}. \citet{trac10} followed a 
procedure similar to that of S10, exploring the thermal and kinetic SZ power
spectra produced for different input parameters of the \citet{bode09}
gas model. 
We adopt the ‘nonthermal20’ model (hereafter the Trac model) presented in
that work, which differs from the S10 simulations by having increased star
formation, lower energy feedback, and the inclusion of 20\% non-thermal
pressure support.
It predicts a significantly smaller value of $D^{\rm tSZ}_{3000}=4.5\,
\mu{\rm K}^2$ when scaled to the SPT $150\,$GHz band.  
The second
template we consider is that produced by \citet{battaglia10} from their
Smoothed-Particle-Hydrodynamics simulations including radiative cooling, star formation and AGN
feedback (hereafter the Battaglia model). This model predicts $D^{\rm tSZ}_{3000}=5.6 \,\mu{\rm K}^2$,  
intermediate between the baseline model and the Trac 
model, and peaks at slightly higher $\ell$ than either of those
models. \citet{shaw10} investigate the impact of cluster astrophysics
on the tSZ power spectrum using halo model calculations in combination
with an analytic model for the ICM. We use the baseline model from
that work (hereafter the Shaw model), which predicts $D^{\rm tSZ}_{3000}=4.7 \,\mu{\rm K}^2$ in the
$150\,$GHz band.  
The model of \citet{shaw10} is also used to rescale
all the model templates as a function of cosmological parameters, as
described in \S\ref{sec:sigma8}.

\begin{figure}
\includegraphics[width=0.5\textwidth]{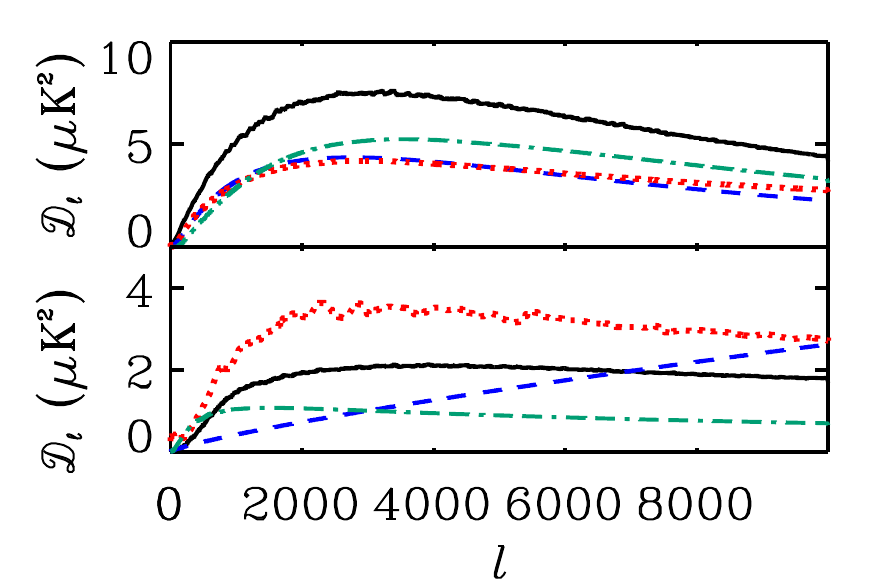}
  \caption[]{Templates used for the tSZ, kSZ, and clustered DSFG power discussed in \S\ref{sec:model}.  The {\em top plot} shows alternate tSZ templates.  The {\bf black, solid line} is the (baseline) S10 model.  The {\bf blue, dashed line} is the Shaw model.  The {\bf red, dotted line} is the Trac model.  The {\bf teal dot-dash line} is the Battaglia model.
The {\em bottom plot} shows both kSZ and clustered DSFG templates.  The {\bf black, solid line} is the (baseline) S10 kSZ model.  The {\bf red, dotted line} is the patchy kSZ model.  The {\bf blue, dashed line} is the (baseline) power-law clustered DSFG template.  The {\bf teal, dot-dash line} is the linear-theory clustered DSFG template.  The clustered DSFG templates have both been normalized to $1\;\mu {\rm K}^2$ at $\ell=3000$.}
\label{fig:templates}
\end{figure}

All four tSZ models exhibit a similar angular scale dependence over the range of multipoles to which SPT is sensitive (see Figure \ref{fig:templates}).  
We allow the normalization of each model to vary in all chains, 
 and detect similar tSZ power in all cases (see Table \ref{tab:tszconstraints}). 
However as we discuss in \S\ref{sec:sigma8}, the difference between models is critical in interpreting the detected tSZ power as a constraint on cluster physics or $\sigma_8$.

\subsubsection{kSZ Power Spectrum}
\label{subsec:kszmodel}

We use the S10 (homogeneous reionization) kSZ template as the baseline kSZ model. 
At $\ell = 3000$, this model predicts $D_{3000}^{kSZ}=2.05\, \mu{\rm K}^2$ of kSZ power.
The kSZ amplitude depends on the details of reionization, and the scaling of the kSZ power with cosmological parameters, particularly $\sigma_8$, is much weaker than
the scaling of the tSZ power.  
We therefore choose to fix the amplitude of the kSZ signal to a model value and allow tSZ to vary independently.  
This treatment differs from \citet{dunkley10}, which uses a single normalization for both SZ components, and \citet{millea10} in which a fitting function is used to calculate the kSZ power as a function of cosmological parameters at each step in the MCMC chain.  
For the current SPT data set, where tSZ and kSZ power are largely degenerate, we expect the differences in 
kSZ treatment to be insignificant.  
This assumption was tested by importance sampling an MCMC chain with variable kSZ amplitude according to the scaling at
$\ell=3000$ described in \citet{millea10}.  As expected, we find no significant difference in fitted parameters.

In \S\ref{sec:params} we will discuss the impact of two alternate kSZ treatments in addition to the baseline model.  
First, we consider a kSZ template that includes the signal from patchy reionization (hereafter the patchy kSZ template).
This template, which was also used in L10, is based upon the ``FFRT'' semi-analytic model of \citet{zahn10}.  
It was calculated in a 1.5 Gpc/h cosmological column where the x- and y-axes correspond to roughly 15 degrees on a side and the z-axis corresponds to redshift, with a median redshift of 8.
The inclusion of patchy reionization increases the kSZ amplitude to at most $3.3 \,\mu{\rm K}^2$, and changes the shape of $D_\ell^{kSZ}$ slightly when compared to the fiducial template (see Figure \ref{fig:templates}).
In the second kSZ treatment, we consider the impact of allowing the overall kSZ amplitude to vary.
In this treatment, we found the small difference between the S10 and patchy kSZ template shapes had no impact on the result.
We therefore use the S10 template for all variable amplitude kSZ constraints.

\subsection{Foregrounds}

At $\ell > 3000$, the measured anisotropy is dominated by
astrophysical foregrounds, particularly bright radio point sources.
We mask all point sources detected at $>\,5\,\sigma$ ($6.4\,$mJy at $150\,$GHz) when estimating the power spectrum.
In order of importance after masking the brightest point sources,  we expect contributions from DSFGs (including their clustering), radio sources
(unclustered), and galactic cirrus.  
We parametrize these galactic and extragalactic foregrounds with an eleven-parameter model.  
Five parameters describe the component amplitudes at $\ell = 3000$: three point source amplitudes in the $150\,$GHz band (clustering and Poisson for DSFGs and Poisson for radio) and a galactic cirrus amplitude in each 
of the two observation bands.   
One parameter describes the correlation between the tSZ and clustered DSFG terms.  
Three parameters describe the effective mean spectral indices of each point source component,
and the final two parameters describe the distribution about the mean
spectral index for the Poisson radio source and DSFG populations.

The frequency dependencies of many of these foregrounds are naturally discussed and modeled in units of flux density (Jy) rather than CMB temperature units.  
This is because the flux density of these foregrounds can be modeled as a power law in frequency, ($S_\nu \propto \nu^{\alpha}$).  
However, in order to model these foregrounds in the power spectrum, we need to determine the ratio of powers in CMB temperature units. 
To calculate the ratio of power in the $\nu_i$ cross $\nu_j$ cross-spectrum to the power at (the arbitrary frequency) $\nu_0$ in units of CMB temperature squared, we multiply the ratio of flux densities by
\begin{equation}
\epsilon_{\nu_1,\nu_2} \equiv \frac{\frac{dB}{dT}|_{\nu_0} \frac{dB}{dT}|_{\nu_0}}{\frac{dB}{dT}|_{\nu_1} \frac{dB}{dT}|_{\nu_2}},
\end{equation}
where $B$ is the CMB blackbody specific intensity evaluated at $T_{\textrm{CMB}}$, and $\nu_i$ and $\nu_j$ are the effective frequencies of the SPT bands.  Note that $\nu_i$ may equal $\nu_j$.
The effective frequencies for each foreground calculated for the measured SPT bandpass (assuming a nominal frequency dependence) are presented in \S\ref{subsec:efffreq}.

\subsubsection{Poisson DSFGs}
\label{subsec:dsfgprior}

Poisson distributed point sources produce a constant ${\it C}_\ell$ (thus, ${\it D}_\ell \propto \ell^2$).
A population of dim ($\lesssim 1\,$mJy) DSFGs is expected to contribute the bulk of the millimeter wavelength point source flux.
This expectation is consistent with recent observations (\citet{lagache07}, \citet{viero09}, H10, \citet{dunkley10}).
Thermal emission from dust grains heated to tens of Kelvin by star light leads to a large effective spectral index ($S_\nu \propto \nu^{\alpha}$), where $\alpha > 2$ between $150$ and $220\,$GHz, for two reasons \citep[e.g.][]{dunne00}.
First, these frequencies are in the Rayleigh-Jeans tail of the blackbody emission for the majority of the emitting DSFGs. 
Second, the emissivity of the dust grains is expected to increase with frequency for dust grains smaller than the photon wavelength (2 mm for 150 GHz) \citep{draine84,gordon95}.

Assuming that each galaxy has a spectral index $\alpha$ drawn from a normal distribution with mean
$\alpha_p$ and variance $\sigma_p^2$, then following \citet{millea10}, the Poisson power 
spectrum of the DSFGs can be written as
\begin{equation}
D^{p}_{\ell,\nu_1,\nu_2} = \amplitudeletter{p} \epsilon_{\nu_1,\nu_2} \eta_{\nu_1,\nu_2} ^{\alpha_{p} + 0.5 {\rm ln}(\eta_{\nu_1,\nu_2}) \sigma_{p}^2} \left( \frac{\ell}{3000}\right)^2
\end{equation}
where $\eta_{\nu_1,\nu_2}  = (\nu_1 \nu_2 / \nu_0^2)$ is the ratio of the frequencies of the spectrum to the base frequency. 
$\amplitudeletter{p}$ is the amplitude of the Poisson DSFG power spectrum at $\ell=3000$ and frequency $\nu_0$.

In principle, the frequency scaling of the Poisson power depends on
both $\alpha_p$ and $\sigma_p$.  
H10 combine the dust spectral energy distributions (SEDs) from \citet{silva98} with an assumed redshift distribution of galaxies to infer  $\sigma_{p} \sim 0.3$ (although they conservatively adopt a prior range of $\sigma_{p} \in [0.2, 0.7]$).  
However, for $\alpha = 3.6$, the ratio of Poisson power at 220 to $150\,$GHz only increases by 2\% for $\sigma_p = 0.3$ versus $\sigma_p = 0$, which is small compared to the constraint provided by the current data.
The complete model includes six parameters describing three measured Poisson powers in two bands.  In order to reduce this to a reasonable number of parameters, we set $\sigma_p=0$ in the baseline model.
The equivalent Poisson radio galaxy spectral variance, $\sigma_{r}$, has a negligible effect on the radio power at $220\,$GHz and is also set to zero.

\subsubsection{Clustered DSFGs}
\label{subsec:clusdsfg}

Because DSFGs trace the mass distribution, they are spatially correlated.
In addition to the Poisson term, this leads to a `clustered' term in the power spectrum of these sources,
\begin{equation}
D^{c}_{\ell,\nu_1,\nu_2} = \amplitudeletter{c} \epsilon_{\nu_1,\nu_2} \eta_{\nu_1,\nu_2} ^{\alpha_{c}} \frac{\modelletter^{c}}{\modelnorm^{c}}.
\end{equation}
$\amplitudeletter{c}$ is the amplitude of the clustering contribution to the DSFG power spectrum at $\ell=3000$ at frequency $\nu_0$.  
The model for the clustering contribution to the DSFG power spectrum is $\modelletter^{c}$, which we have divided by $\modelnorm^{c}$ to make a unitless template which is normalized at $\ell = 3000$.  

We consider two different shapes for the angular power spectrum
due to the clustering of the DSFGs, shown in Figure~\ref{fig:templates}.
The first is the shape of the fiducial model used in Hall et al. (2010).  
This spectrum was calculated assuming that light is a biased tracer of mass fluctuations calculated in linear perturbation theory.  
Further, the model assumes all galaxies have the same SED, that of a grey body, and the redshift distribution of the luminosity density was given by a parametrized form, with parameters adjusted to fit SPT, BLAST, and Spitzer power spectrum measurements.  Here we refer to this power spectrum shape as the ``linear theory'' template.

On the physical scales of relevance, galaxy correlation functions are surprisingly well-approximated by a power law.  
The second shape we consider is thus a phenomenologically-motivated power law with index chosen to match the correlation properties observed for Lyman-break galaxies at $z \sim 3$ \citep{giavalisco98,scott99}.  We adopt this ``power law'' model in the baseline model, and use a template of the form $\modelletter^{c} \propto \ell^{0.8}$ in all cases except as noted.

We also consider a linear combination of the linear-theory shape and the power-law shape.  
Such a combination can be thought of as a rough approximation to the two-halo and one-halo terms of a halo model.  The present data set has limited power to constrain the ratio of the two terms; however, we do find a slight preference for the pure power-law model over the pure linear-theory model, as discussed in sections \ref{subsec:twohalo} and \ref{subsec:1halo2halo}.
 
The spectrum of the clustered component should be closely related to the spectrum of the Poisson component 
since they arise from the same sources (H10). 
However, the clustered DSFG term is expected to be more significant at higher redshifts, where the rest-frame frequency of the SPT observing bands is shifted to higher frequencies and out of the Raleigh-Jeans approximation for a larger fraction of the galaxies.
One might therefore anticipate a slightly lower spectral index for the clustered term.

We assume a single spectral index for the Poisson and clustered DSFG components in the baseline model ($\alpha_{c} = \alpha_{p}$), and
explore the impact of allowing the two spectral indices to vary independently in \S\ref{subsec:freealphac}.  
In general, we require $\alpha_{c} = \alpha_p,$ except where otherwise noted.

\subsubsection{tSZ-DSFG correlation}
\label{subsec:tszdsfgcorrel}

The tSZ signal from galaxy clusters and the clustered DSFG signal are both biased tracers of the same dark matter distribution, although the populations have different redshift distributions.
Thus, it is natural to expect these two signals to be partially correlated 
and this correlation may evolve with redshift and halo mass.
At frequencies below the tSZ null, this should result in spatially anti-correlated power.
At $150\,$GHz, some models that attempt to associate emission with individual cluster member galaxies predict 
anti-correlations of up to tens of percent (S10).
L10 argue against this being significant in the SPT $150\,$GHz band, because cluster members observed at low
redshift have significantly less dust emission than field galaxies.
It is, of course, possible for this emission to be significant at higher redshifts where much of the
tSZ power originates.
Current observations place only weak constraints on this correlation.
(As we show in \S\ref{subsec:dusty-tsz-correl}, the data presented here are consistent with an anti-correlation coefficient ranging from zero to tens of percent.)
To the extent that the Poisson and clustered power spectra have the same spectral index, the differenced band 
powers discussed in \S\ref{sec:l10s10comp} (as well as in L10) are insensitive to this correlation.

A correlation, $\gamma(\ell)$, between the tSZ and clustered DSFG signals introduces an additional term in the model,
\begin{equation}\begin{split}
&D^{tSZ-DSFG}_{\ell,\nu_1,\nu_2}  =  \\
&\,\,\,\, \gamma(\ell)  \left(  \sqrt{D^{c}_{\ell,\nu_1,\nu_1}  D^{tSZ}_{\ell,\nu_2,\nu_2}}  + \sqrt{D^{c}_{\ell,\nu_2,\nu_2}  D^{tSZ}_{\ell,\nu_1,\nu_1} } \right).
\end{split}
\end{equation}
We will assume no tSZ-DSFG correlation in all fits unless otherwise noted.
In \S\ref{subsec:dusty-tsz-correl}, we will investigate constraints on the degree of correlation 
and the impact on other parameters from allowing a DSFG-tSZ correlation.

\subsubsection{Radio galaxies}

The brightest point sources in the SPT maps are coincident with known radio sources and have spectral indices consistent with synchrotron emission \citep[hereafter V10]{vieira10}. 
Much like the Poisson DSFG term, the model for the Poisson radio term can be written as
\begin{equation}
D^{r}_{\ell,\nu_1,\nu_2} = \amplitudeletter{r} \epsilon_{\nu_1,\nu_2} \eta_{\nu_1,\nu_2} ^{\alpha_{r} + 0.5 {\rm ln}(\eta_{\nu_1,\nu_2}) \sigma_{r}^2} \left( \frac{\ell}{3000}\right)^2.
\end{equation}
The variables have the same meanings as for the treatment of the DSFG Poisson term.
Clustering of the radio galaxies is negligible.

In contrast to DSFGs, for which the number counts climb steeply towards lower flux, S$^2$dN/dS for synchrotron sources is fairly constant. 
As a result, the residual radio source power ($\amplitudeletter{r}$) in the map depends approximately linearly on the flux above which discrete sources are masked. 
For instance, if all radio sources above a signal to noise threshold of $5\,\sigma$ ($6.4\,$mJy) at $150\,$GHz are masked, the \citet{dezotti05} source count model predicts a residual radio power of $\amplitudeletter{r} = 1.28\,\mu{\rm K^2}$.
For a $10\,\sigma$ threshold, the residual radio power would be $\amplitudeletter{r} = 2.59\,\mu{\rm K}^2$.

The de Zotti model is based on radio surveys such as NVSS extrapolated to $150\,$GHz using a set spectral index for each subpopulation of radio sources. 
Extrapolating over such a large frequency range obviously introduces significant uncertainties; however, we do have an important 
cross-check on the modeling from the number counts of bright radio sources detected in the SPT or 
ACT maps (V10, \citet{marriage10}).
The de Zotti number counts appear to over-predict the number of very high flux radio sources reported by both experiments.
We find that the de Zotti model should be multiplied by $0.67\pm 0.14$ to match the number of $>20\,$mJy radio sources in the SPT source catalog (V10).
There are relatively few radio sources at these fluxes, and V10 and \citet{marriage10} have partially overlapping sky coverage.
The counting statistics on these relatively rare objects will be improved significantly with the full SPT survey.
The discrepancy disappears for lower flux radio sources and we find that, for 
$6.4\,{\rm mJy} < S_{150} <20\,$mJy  $150\,$GHz, the best fit normalization of the de Zotti model to fit 
the radio sources in the V10 catalog is $1.05\pm 0.14$. 
Throughout this work, we use the de Zotti model prediction with a 15\% uncertainty as a prior for residual power from sources below $6.4\,$mJy.
Relaxing this radio prior by expanding the $1\,\sigma$ amplitude prior about the \citet{dezotti05} model from 15\% to 50\% has no significant impact on resulting parameter constraints.

We analyze the two-frequency V10 catalog to determine the radio population mean spectral index ($\alpha_r$) and scatter in spectral indices ($\sigma_r$) at these frequencies.
For each source, we take the full (non-Gaussian) likelihood distribution for the source spectral indices
from V10.
We combine all sources above $5\,\sigma$ at $150\,$GHz with a probability of having a synchrotron-like spectrum of at least 50\% in order to calculate the likelihoods of a given $\alpha_r$ and $\sigma_r$.
(We note that the results are unaffected by the specific choice of threshold probability for being a synchrotron source.) 
The best-fit mean spectral index is $\alpha_r=-0.53\pm0.09$ with a $2\,\sigma$ upper limit on the intrinsic scatter of $\sigma_r < 0.3$.
With the present power spectrum data, both this uncertainty and intrinsic scatter are undetectable.
We have checked this assumption by allowing the spectral index to vary with a uniform prior between $\alpha_{r} \in [-0.66, -0.21]$ instead of fixing it to the best-fit value of -0.53 for several specific cases, and found no significant difference in the resulting parameter constraints.
We therefore assume $\alpha_r = -0.53$ and $\sigma_r=0$ for all cases considered.

\subsubsection{Galactic cirrus}
\label{subsec:cirrusprior}

In addition to the foregrounds discussed above, we include a fixed galactic cirrus term parametrized as
\begin{equation}
D^{cir}_{\ell,\nu_1,\nu_2} = \amplitudeletter{cir, \nu_1,\nu_2} \left( \frac{\ell}{3000}\right)^{-1.5}.
\end{equation}
H10 measured the Galactic cirrus contribution in the \five\ field by cross-correlating the SPT map and model eight of \citet[hereafter FDS]{finkbeiner99} for the same field.
We extend that analysis to the combined data set by scaling the \five\ cirrus estimate by the ratio of the auto-correlation power in the FDS map of each field.
We find the \twentythree\ field has 15\% of the cirrus power measured in the \five\ field.
As the fields are combined with equal weight, we expect the effective cirrus amplitude in the combined bandpowers to be 0.575 times the \five\ amplitudes reported above.
Based on this analysis, we fix the amplitude of the cirrus term, ${\it D}_\ell \propto \ell^{-1.5}$, in all parameter fits to be $(\amplitudeletter{cir, 150,150},  \amplitudeletter{cir, 150,220}, \amplitudeletter{cir, 220,220}) =(0.35, 0.86, 2.11)\,\mu{\rm K}^2$ respectively.
Note that the cirrus power has a dust-like spectrum and is effectively removed in the differenced bandpowers discussed in \S\ref{sec:l10s10comp}.
We assume there is zero residual cirrus power in the DSFG-subtracted bandpower analyses. 
Although we include this cirrus prior in the multi-frequency fitting, it has no significant impact on results; marginalizing over the amplitudes in the two bands or 
setting them to either extreme (zero or the high value in the \five\ field) does not change the resulting parameter values.

\subsection{Effective frequencies of the SPT bands}
\label{subsec:efffreq}

Throughout this work, we refer to the SPT band center frequencies as 150 and 220 GHz.
The data are calibrated to CMB temperature units, hence the effective frequency is irrelevant for a CMB-like source.
However, the effective band center will depend (weakly) on the source spectrum for other sources.
The actual SPT spectral bandpasses are measured by Fourier transform spectroscopy, and the effective band center frequency is calculated for each source, assuming a nominal frequency dependence.
For an $\alpha = -0.5$ (radio-like) source spectrum, we find band centers of 151.7 and $219.3\,$GHz.
For an $\alpha = 3.8$ (dust-like) source spectrum, we find band centers of 154.2 and $221.4\,$GHz.
The band centers in both cases drop by $\sim$$\,0.2\,$GHz if we reduce $\alpha$ by 0.3.
We use these radio-like and dust-like band centers to calculate the frequency scaling between the SPT bands for the radio and DSFG terms.
For a tSZ spectrum, we find band centers of 153.0 and $220.2\,$GHz.
The 220 band center is effectively at the tSZ null; the ratio of tSZ power at 220 to $150\,$GHz is 0.2\%.

\section{Parameter results}
\label{sec:params}

\subsection{Baseline model results}
\label{sec:baseline}

\begin{table*}[ht!]
\begin{center}
\caption{\label{tab:params01} Parameters from multi-frequency fits}
\small
\begin{tabular}{lccccccccc}
\hline
model& $\amplitudeletter{tSZ}$& $\amplitudeletter{kSZ}$& $\amplitudeletter{P}$&
$\amplitudeletter{C}$& $\alpha_P$& $\alpha_C$& $\amplitudeletter{P,220}$&
$\amplitudeletter{C,220}$ \\
\hline
Baseline model & $3.5 \pm 1.0$ & $[2.0] $ & $7.4 \pm 0.6$ & $6.1 \pm 0.8$ & $3.58 \pm 0.09$ & $-- $ & $70 \pm 4$ & $57 \pm 8$ \\
Patchy kSZ & $2.9 \pm 1.0$ & $[3.3] $ & $7.4 \pm 0.6$ & $5.7 \pm 0.8$ & $3.62 \pm 0.09$ & $-- $ & $72 \pm 4$ & $55 \pm 8$ \\
Free kSZ & $3.2 \pm 1.3$ & $2.4 \pm 2.0$ & $7.4 \pm 0.6$ & $5.9 \pm 1.0$ & $3.59 \pm 0.11$ & $-- $ & $71 \pm 5$ & $57 \pm 9$ \\
Free $\alpha_{c}$ & $3.4 \pm 1.0$ & $[2.0] $ & $7.9 \pm 1.0$ & $5.4 \pm 1.4$ & $3.46 \pm 0.21$ & $3.79 \pm 0.37$ & $69 \pm 5$ & $59 \pm 8$ \\
Linear-theory clustering & $3.5 \pm 1.0$ & $[2.0] $ & $9.3 \pm 0.5$ & $4.9 \pm 0.7$ & $3.59 \pm 0.09$ & $-- $ & $89 \pm 5$ & $47 \pm 7$ \\
Free $\alpha_{c}$ \& linear theory & $3.4 \pm 1.0$ & $[2.0] $ & $9.2 \pm 0.7$ & $5.3 \pm 1.4$ & $3.62 \pm 0.14$ & $3.47 \pm 0.34$ & $90 \pm 5$ & $46 \pm 7$ \\
tSZ-DSFG correlation & $3.4 \pm 1.0$ & $[2.0] $ & $9.3 \pm 0.7$ & $4.9 \pm 0.8$ & $3.60 \pm 0.13$ & $-- $ & $89 \pm 5$ & $46 \pm 7$ \\
tSZ-DSFG cor, free kSZ & $1.8 \pm 1.4$ & $5.5 \pm 3.0$ & $9.1 \pm 0.7$ & $4.4 \pm 0.9$ & $3.62 \pm 0.14$ & $-- $ & $89 \pm 5$ & $43 \pm 8$ \\

\\
\multicolumn{7}{c}{Priors and Templates} \\
\hline
&&&&&&DG clustered&\\
model & $N_{pars}$ & Free $\alpha_{c}$ & Free $\amplitudeletter{\rm kSZ}$ & tSZ model & kSZ model &  model &  $-2\Delta {\rm ln} \mathcal{L}$\\
\hline
Baseline model & 10 & No & No & S10 & S10 & power law & -- \\
Patchy kSZ & 10 & No & No & S10 &Patchy & power law & $1$\\
Free kSZ & 11 & No &Yes &  S10 &S10 & power law & $0$\\
Free $\alpha_{c}$ & 11 & Yes & No & S10 &S10 & power law & $0$\\
Linear-theory clustering  &  10 & No & No & S10 &S10 & linear theory & $4$ \\
Free $\alpha_{c}$ \& linear theory& 11 & Yes & No & S10 &S10 & linear theory & $4$ \\
tSZ-DSFG cor & 11 & No & No & S10 & S10 & linear theory & 4 \\
tSZ-DSFG cor, free kSZ & 12 & No & Yes & S10 & S10 & linear theory & 4\\

\end{tabular}
\vspace{1em}
\tablecomments{Parameter constraints from multi-frequency fits to the models described in \S\ref{sec:model}.  
The amplitudes are $\mathcal{D}_\ell$ at $\ell=3000$ for the $150\,$GHz band, in units of $\mu{\rm K}^2$.
  In all cases, the error specified is one half of the 68\% probability width for a given parameter after marginalizing over other parameters as described in the text.  Where $\alpha_{c}$ is unspecified, both the clustered and Poisson DSFG components share an identical spectral index given by $\alpha_{p}$.
The majority of the chains use a fixed kSZ spectrum; in these cases, the kSZ value appears in brackets. 
In addition to the model parameters described in \S\ref{sec:model}, we include two derived parameters, $\amplitudeletter{P,220}$  and $\amplitudeletter{C,220}$, which are the amplitude of the Poisson and clustered DSFG component in the SPT 220 GHz band, respectively.
We also summarize the free parameters, templates, and relative goodness of fit for each case.
}
\end{center}
\end{table*}

\begin{table}[ht!]
\begin{center}
\caption{\label{tab:deltachisq} $\Delta \chi^2$ with the addition of model components}
\small
\begin{tabular}{lcc}
\hline
Variable & dof &$-2\Delta {\rm ln} \mathcal{L}$\\
\hline
point sources (Poisson)  & 2 & -7821\\
point sources (including clustering)  & 1 & -104 \\
tSZ & 1 & -25\\
kSZ & 0 & -1\\

\end{tabular}
\vspace{1em}
\tablecomments{The addition of DSFG and SZ terms in the model significantly improves the quality of the maximum likelihood fit. 
From top to bottom, the improvement in the fit $\chi^2$ as each term is added to a model including the primary CMB anisotropy and all the terms above it, e.g., the tSZ row shows the improvement from adding the tSZ effect with a free amplitude to a model including the CMB,  and point source power, including the contribution from clustering of the DSFGs.
The second column (dof), lists the number of degrees of freedom in each component. 
The free Poisson point source parameters are the amplitude and spectral index of the DSFGs.
Note that the third Poisson component (the radio power) is constrained by an external prior rather than the data and we have marginalized over the allowed radio power.
The free clustered point source parameter is the amplitude of the power-law clustered template. 
The spectral index of the DSFG clustering term is fixed to that of the Poisson term.
The last two rows mark the addition of the S10 tSZ template with a free amplitude, and the addition of the S10 homogeneous kSZ template with a fixed amplitude.
We do not find a significant improvement in the likelihood with any extension to this model, including allowing the kSZ amplitude to vary, or allowing independent spectral indices for the two DSFG components.
}
\end{center}
\end{table}

As discussed in \S\ref{sec:model}, 
the baseline secondary anisotropy and foreground model contains four free parameters (beyond the six $\Lambda$CDM parameters): the tSZ amplitude, the Poisson DSFG amplitude, the clustered DSFG amplitude, and the DSFG spectral index.  
For the tSZ component, we use the S10 tSZ model scaled by a single amplitude parameter.  
For the clustered DSFG component, we use the power-law model presented in \S\ref{subsec:clusdsfg}, again scaled by an amplitude parameter. 
As discussed in \S\ref{subsec:1halo2halo}, the data show a slight preference for the power-law clustered 
point source shape model compared to the linear-theory model that was adopted in L10.
We require that both the clustered and Poisson DSFG components have the same spectral index, which we model as the free parameter $\alpha_{p},$ with no intrinsic scatter ($\sigma_{p}=0).$
Table~\ref{tab:deltachisq} shows the improvement in the quality of the fits with the sequential
introduction of free parameters to the original $\Lambda$CDM primary CMB model.
Beyond this baseline model, adding additional free parameters does not improve the quality of the fits. 

The additional parameters of the model discussed in \S\ref{sec:model} are all fixed or constrained 
by priors.  
The kSZ power is fixed to the $D^{\rm kSZ}_{3000}=2.05\,\mu{\rm K}^2$ expected  
for the homogeneous reionization model of \citet{sehgal10}.
Radio galaxies are included with a fixed spectral index $\alpha_{r}=-0.53$ and zero 
intrinsic scatter 
and an amplitude $D^r_{3000}=1.28\,\mu{\rm K}^2$ with a 15\% uncertainty.
Here, and throughout this work, we also include a constant Galactic cirrus power as described in 
\S\ref{subsec:cirrusprior}.
In addition to astrophysical model components, for each band we include a term to adjust the overall 
temperature calibration and three terms to parametrize the uncertainty in the measurement of the 
beam function. 
This baseline model provides the most concise interpretation of the data.
As will be demonstrated in subsequent sections, relaxing priors on the fixed or tightly constrained
parameters in this model has minimal impact on the measured tSZ power.  

\begin{figure*}[ht]\centering

\includegraphics[width=0.7\textwidth]{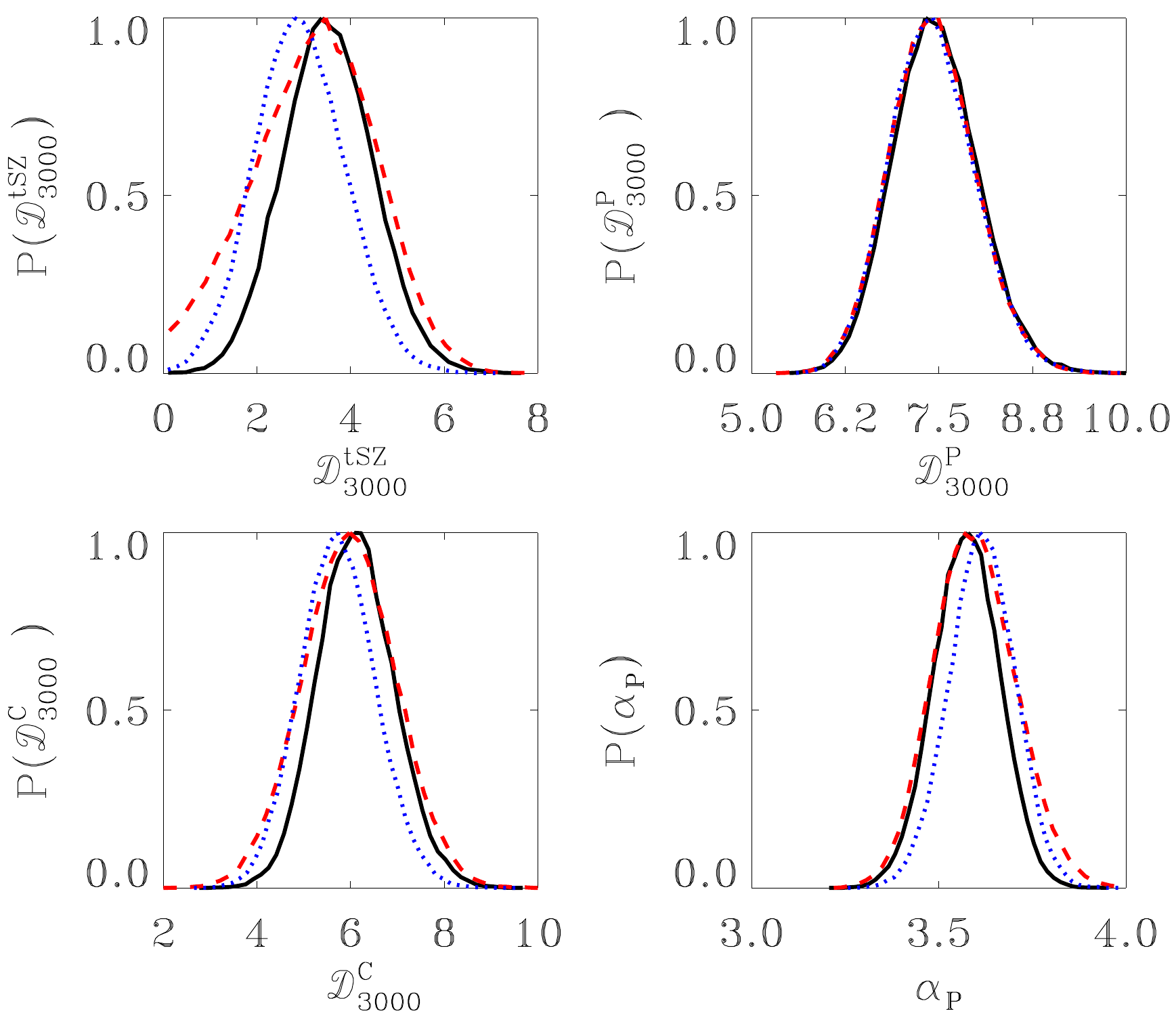}
  \caption[]{1D likelihood plots for model parameters for three different kSZ treatments.  The {\bf black, solid line} is the baseline (fixed amplitude, S10 homogeneous model) kSZ model.  The {\bf blue dotted line} is the patchy kSZ model with fixed amplitude.  The {\bf red, dashed line} is the model where the amplitude of the (default) kSZ template is free to vary.  All powers are measured at  $\ell = 3000$ and $150\,$GHz. 
  The {\it top left panel} shows constraints on the tSZ power spectrum. 
  The preferred tSZ power drops as the assumed kSZ power increases; this trade-off can be seen in more detail in Figure \ref{fig:2dvaryksz}.
  The {\it top right panel} shows the likelihood curves for the Poisson DSFG power,
    and the {\it bottom left panel} shows the same for the clustered DSFG power.
    The {\it bottom right panel} shows the Poisson DSFG spectral index.
    There is a slight decrease in clustered DSFG power with the increased kSZ power expected from patchy reionization; this is accompanied by small shift to higher spectral indices, preserving the higher-frequency bandpowers.
  }
\label{fig:1dvaryksz}
\end{figure*}

Parameter constraints from this baseline model are listed in Table \ref{tab:params01} and Figure \ref{fig:1dvaryksz}.
For the tSZ power, we find $D_{3000}^{tSZ}=3.5 \pm 1.0 \,\mu{\rm K}^2$, which is consistent with previous measurements (L10, \citet{dunkley10}) and significantly lower than predicted by the S10 model.
We detect both the Poisson and clustered DSFG power with high significance, with amplitudes of $D^{p}_{3000}=7.4 \pm 0.6\, \mu {\rm K}^2$ and $D^{c}_{3000}=6.1 \pm 0.8 \,\mu{\rm K}^2$ respectively.
The shared spectral index of these components is found to be $\alpha=3.58 \pm 0.09$, in line with both theoretical expectations and previous work (H10, \citet{dunkley10}).
The amplitude of the Poisson term is significantly lower than what was measured in H10;
however, as we show in the following sections, this discrepancy is eliminated by adopting the linear-theory clustered DSFG template used in that work.

We explore the effect of altering this baseline model by replacing the tSZ template with the three alternatives discussed in \S\ref{sec:tszmodels}: the Shaw, Trac, and Battaglia models.
As anticipated, we find no significant difference between the models for any of the measured parameters.  
The resulting tSZ powers are listed in Table \ref{tab:tszconstraints};  they are statistically identical.  
The other model parameters also vary by less than 1\% for all four tSZ model templates.

\begin{figure*}[ht]\centering
\includegraphics[width=0.9\textwidth]{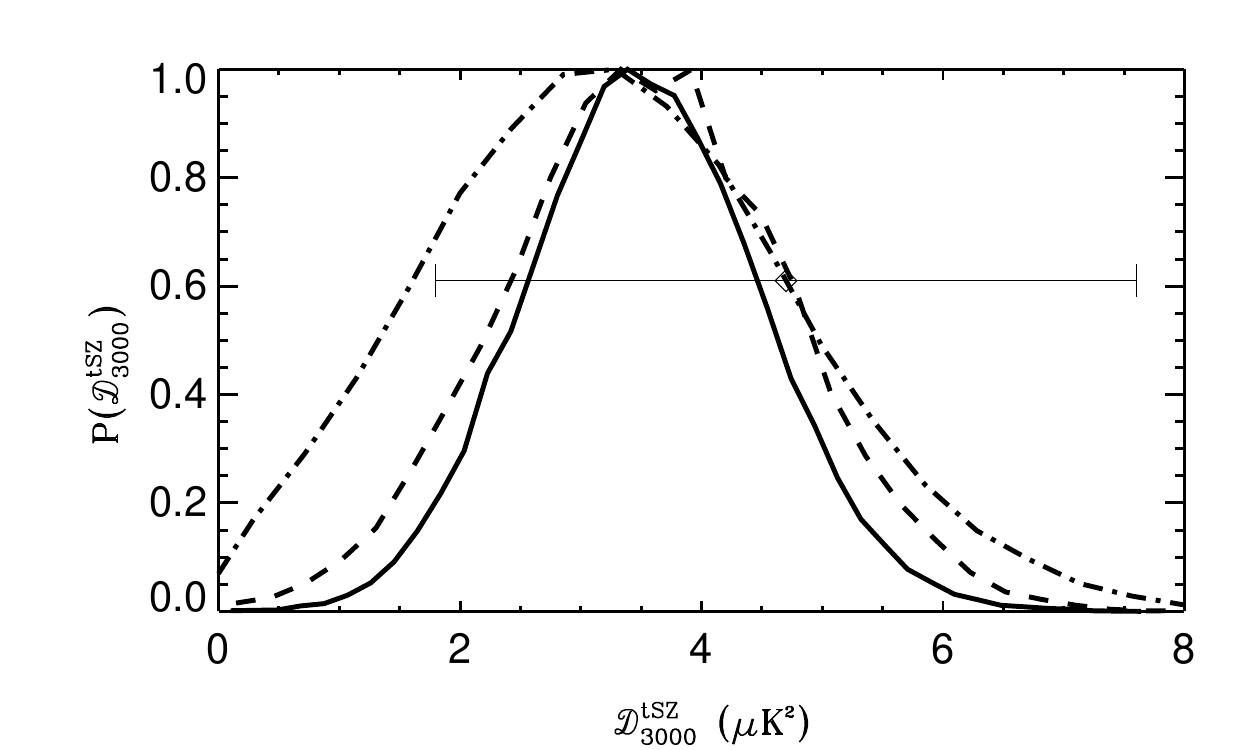} 
    \caption[]{The {\bf solid black bline} shows the 1D likelihood surface for $D_{3000}^{tSZ}$ from the multi-frequency analysis presented in \S\ref{sec:baseline} .  The baseline model with the S10 homogeneous kSZ template has been assumed.  Also plotted are the equivalent results from L10 ({\bf dot-dash line}) and the DSFG-subtracted analysis in \S\ref{sec:l10s10comp} of this work ({\bf  dashed line}).   In addition, the median value and 68\% confidence region for the tSZ amplitude from \citet{dunkley10} is shown by the {\bf diamond and thin horizontal line}.
  }
\label{fig:aszlike1d}
\end{figure*}

\subsection{kSZ variants} 
\label{sec:kSZ}

The baseline kSZ model is the homogeneous reionization model presented in S10.
In this section, two alternative kSZ cases are considered.
First, for the same set of model parameters, we change the (fixed) kSZ power spectrum from the baseline homogeneous reionization model to a template with additional power from patchy reionization.
In the second case, we allow the kSZ amplitude to vary in order to explore the tSZ-kSZ degeneracy and to set upper limits on the SZ terms.

We explore the effect of replacing the homogeneous kSZ template with the patchy kSZ template described in \S\ref{subsec:kszmodel}.
The patchy kSZ model increases the power at $\ell = 3000$ by 50\% to $D^{kSZ}_{3000}=3.25\,\mu{\rm K}^2$, and has a slightly modified angular scale dependence with additional large-scale power.
The likelihood of the best-fit model is nearly unchanged from the homogeneous kSZ case.
The increase in kSZ power by $1.2 \,\mu{\rm K}^2$ leads to a decrease in tSZ power by $0.6 \,\mu{\rm K}^2$, as one would expect given the near-degeneracy along the tSZ + 0.5 kSZ axis which we discuss below.
The kSZ spectrum shape difference has no impact on the results.

\begin{figure*}[ht]\centering

\includegraphics[width=0.5\textwidth]{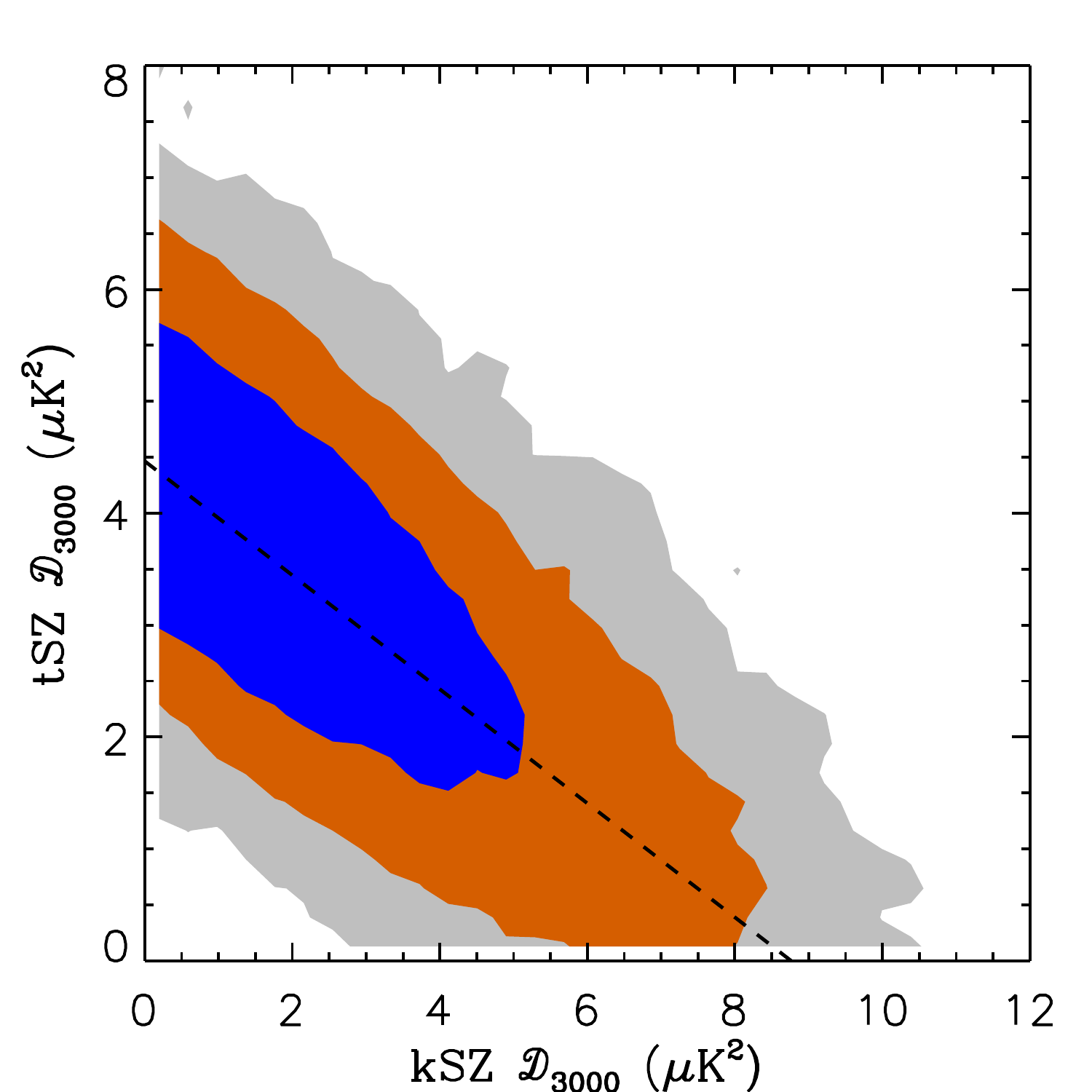}
  \caption[]{2D likelihood contours for $D_{3000}^{\rm tSZ}$ vs. $D_{3000}^{\rm kSZ}$. 
  The 68\%, 95\% and 99.7\% contours are marked with blue, orange, and gray respectively.
  Both tSZ and kSZ model templates are from S10.
  As in the differenced spectrum analysis, the data primarily constrain a linear combination $D_{3000}^{tSZ} + 0.5 D_{3000}^{kSZ} = 4.5\pm 1.0\,\mu{\rm K}^2$.
The dashed line shows this axis of near degeneracy.
  Due to this degeneracy, neither parameter is significantly detected above zero. 
However, 95\% CL upper limits of 5.3 and $6.5\,\mu{\rm K}^2$ can be placed on the tSZ and kSZ respectively.

  }
\label{fig:2dvaryksz}
\end{figure*}

As an alternative to the fixed kSZ models discussed above, we also allow the kSZ amplitude to vary, adding one additional parameter to MCMC chains.
With two frequencies, the current SPT data does not allow the separation of the tSZ, kSZ, and clustered point source components.
The chains show a strong degeneracy between the components, as shown in Figure \ref{fig:2dvaryksz}.
The best-constrained eigenvector in the kSZ/tSZ plane is approximately tSZ + 0.5 kSZ, very similar to the proportionality used in the differenced bandpowers in \S\ref{sec:l10s10comp}.
For this linear combination, we find  $D^{tSZ}_{3000} + 0.5\,D^{kSZ}_{3000} = 4.5\pm 1.0 \,\mu{\rm K}^2$.
We note that the data show some preference for a positive tSZ power, at just under $2\sigma.$
We tested whether this preference is due to the detailed shapes of the two templates by swapping the tSZ and kSZ 
templates.
This model swap had no significant impact on the 2d likelihood surface, indicating that the preference for positive tSZ power is driven by the unique frequency signature of the tSZ.
When marginalizing over the tSZ amplitude and other parameters, 
we find the upper limit on the kSZ power to be $D^{kSZ}_{3000} < 6.5\,\mu{\rm K}^2$ at 95\% confidence.
The corresponding upper limit on tSZ power is $D^{tSZ}_{3000} < 5.3\,\mu{\rm K}^2$ at 95\% confidence.

\subsection{Clustered DSFG extensions to the baseline model}
\label{sec:varydsfg}

\begin{figure*}[ht]\centering
\includegraphics[width=0.7\textwidth]{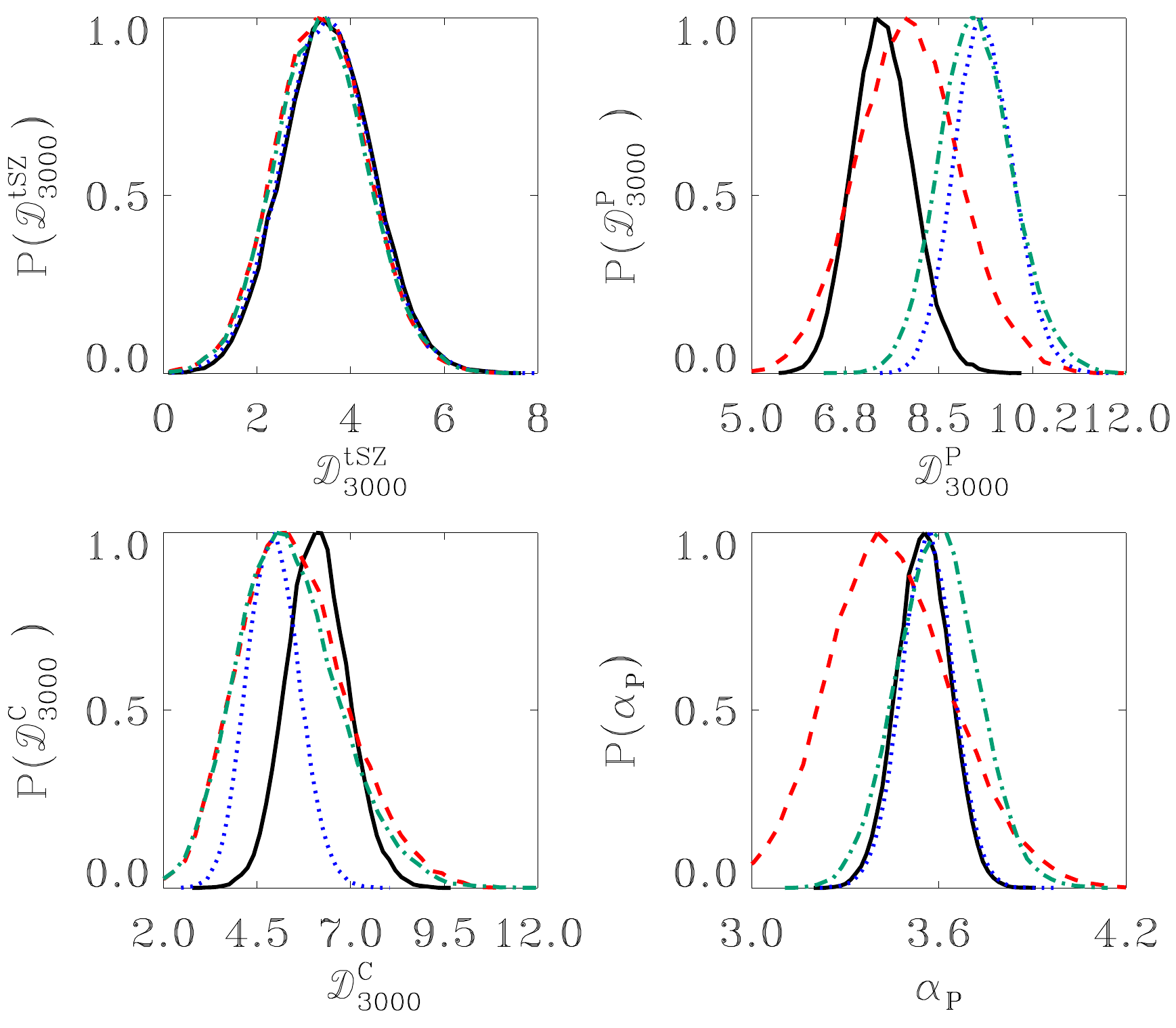}
  \caption[]{
1D likelihood plots of model parameters for different DSFG clustering treatments. 
Clockwise from top left: the 1D likelihood curves for the amplitudes of the tSZ effect, Poisson DSFG amplitude, Poisson DSFG spectral index and clustered DSFG amplitude.
Four cases are plotted for each parameter: 
The {\bf black solid line} indicates the baseline model, where $\alpha_{c}=\alpha_{p}$ and the power-law clustered template is used.
The {\bf red dashed line} indicates the case where $\alpha_{c}$ is a free parameter and the power-law clustered template is used.
The {\bf blue dotted line} indicates the case where $\alpha_{c}=\alpha_{p}$ and the linear-theory clustered template is used.
The {\bf blue-green dot-dash line} indicates the case where $\alpha_{c}$ and $\alpha_{p}$ are independent and the linear-theory clustered template is used.
Using the power-law clustered template tends to reduce the power in the Poisson component compared to the linear-theory template.  
This is expected as the power-law shape has more power at high-$\ell$ than the linear-theory template and the Poisson component also increases with $\ell$ ($D_\ell \propto \ell^2$). 
Including an additional free parameter ($\alpha_{c}$) tends to increase the uncertainty on DSFG-related parameters but not on tSZ.
  }
\label{fig:dsfglike1d}
\end{figure*}

The baseline model contains DSFG sources that generate both Poisson and clustering contributions to the
power spectrum.  The spectral index of the Poisson power and clustering power is fixed to be the same.
In this section, we examine two changes to this baseline model. 
First, we allow the clustered DSFG spectral index to vary independently, 
and second, we change the clustered template from a power-law to linear-theory model. 
We explore all four permutations of these two Boolean cases (one of the four is the baseline case).
One dimensional likelihood curves for each case are shown in Figure~\ref{fig:dsfglike1d} and discussed below.
We also consider a mixture of the power-law and linear-theory templates, each with its own variable amplitude, again for a total of five free parameters. 
As we will discuss in \S\ref{subsec:1halo2halo}, the SPT data slightly prefer the power-law template.

\subsubsection{Free $\alpha_{c}$}
\label{subsec:freealphac}

As discussed in \S\ref{subsec:dsfgprior}, the Poisson and clustered DSFG amplitudes are sensitive to different weightings of the DSFGs with redshift and brightness and thus we expect these components to have slightly offset spectral indices.
For instance, the \citet{righi08} model predicts $\alpha_p -  \alpha_c = 0.2$.
To study this offset, we allow the two spectral indices to vary independently. 
The resulting spectral indices for the clustered and Poisson DSFG terms separate by approximately one $\sigma$;
however, this shift is in the opposite direction from what the \citet{righi08} model predicts. 
With the addition of this additional free parameter, the uncertainty on the Poisson and clustered 
amplitudes and spectral indices increase significantly. 
At $220\,$GHz, both DSFG powers are essentially unchanged, while the $150\,$GHz 
Poisson and clustered amplitudes slightly increase and decrease, respectively.

\subsubsection{Linear-theory model}
\label{subsec:twohalo}

We adopt the H10 model as the linear-theory clustered DSFG model. 
The angular scale dependence of this model is very similar to the tSZ template and nearly flat in $D_\ell$.
Swapping between the power-law and linear-theory models does not change the measured tSZ power.
However, it does shift some power between the Poisson and clustered DSFG terms. 
The angular scale dependence of the power-law model is intermediate between the linear-theory model and Poisson terms.
As a result, the flatter, linear-theory model leads to more Poisson and less clustered power.
At $\ell = 3000$ and $150\,$GHz, the Poisson DSFG power increases by $3\,\sigma$ to $D^p_{3000}=9.3 \pm 0.5 \,\mu{\rm K}^2$ from $7.4 \pm 0.6 \,\mu{\rm K}^2$.
The clustered DSFG power decreases correspondingly by $1\,\sigma$ to $D^c_{3000}=4.9 \pm 0.7 \,\mu{\rm K}^2$.
Note that the DSFG spectral index has dropped compared to what was measured in H10 from $3.86 \pm 0.23$ to $\alpha_{p}=3.59 \pm 0.09$.
The change in spectral index is driven by the $220\,$GHz power: the Poisson power is $2\,\sigma$ lower and the clustered power is $1\,\sigma$ lower than what was measured in H10.

\subsubsection{Free $\alpha_{c}$ \& linear-theory model}

Finally, we consider the case where we allow $\alpha_{c}$ as a free parameter with the 
linear-theory clustered DSFG template.
The resulting chains include features from both the free-$\alpha_{c}$ and linear-theory cases: 
the Poisson power increases and tSZ power decreases, while uncertainties on the clustered amplitude and all spectral indices increase.
The spectral indices of the clustered and Poisson DSFG terms do not separate as much as they do in the power-law, free-$\alpha_c$ chain.  

\subsubsection{Discriminating between the power-law and linear-theory models}
\label{subsec:1halo2halo}

In principle, the likelihood of a fit to the data can be used to distinguish between the power-law and linear-theory clustered DSFG templates described above.
We explore this in two ways.
First we consider the improvement in the best-fit log-likelihood between the models which use each template. 
The power-law model produces a marginally better fit to the data, with $\Delta {\rm ln} \mathcal{L} = -2$ when compared to the linear-theory model.

\begin{figure}
  \includegraphics[width=0.5\textwidth]{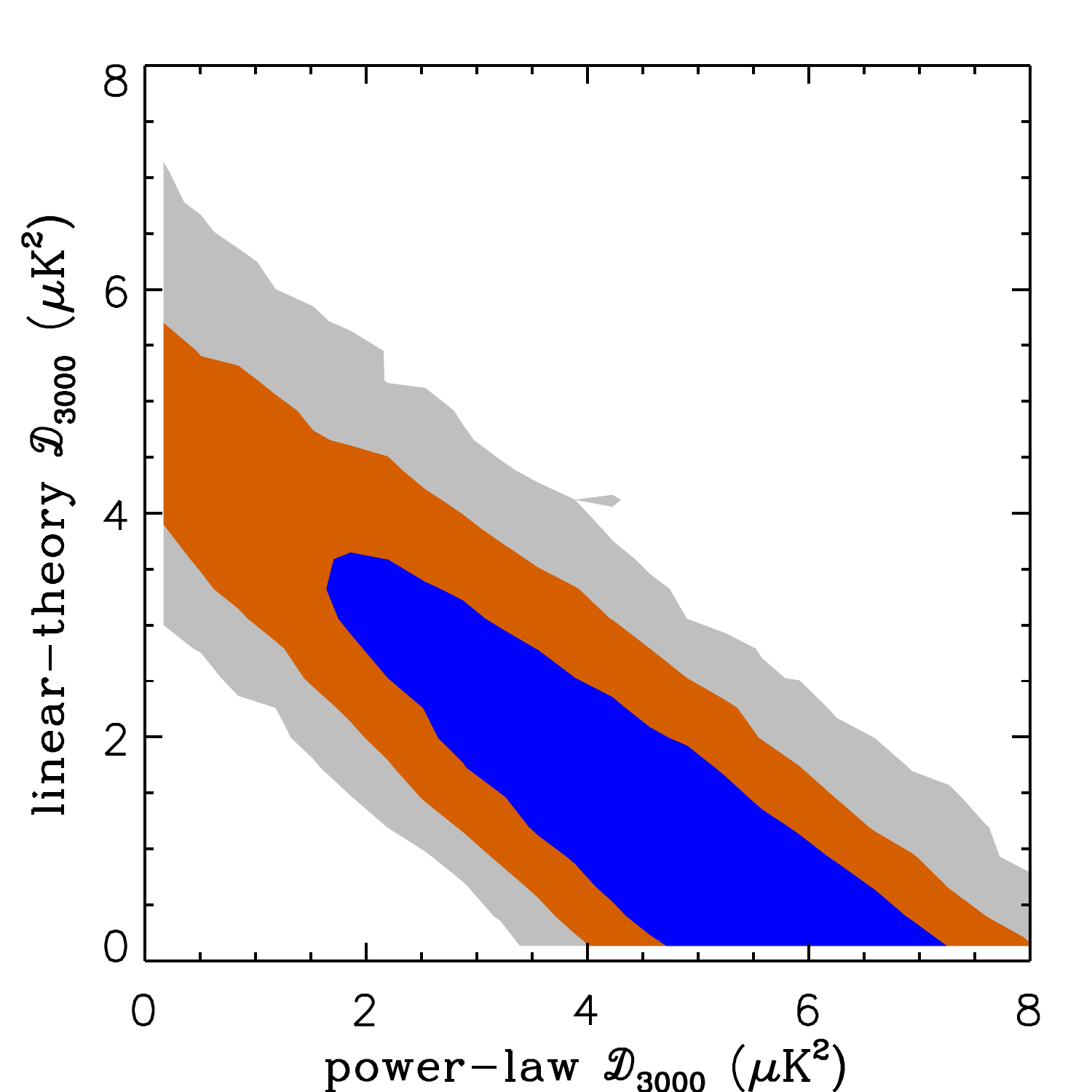}
  \caption[]{68\% ({\bf blue}), 95\% ({\bf orange}) and 99.7\% ({\bf grey}) likelihood contours for the amplitudes of clustered DSFG components for a model containing both power-law and linear-theory components with independently variable amplitudes.  The power-law model is favored at the $1\,\sigma$ level; the data do not clearly discriminate between the power-law shape, linear-theory shape, or a linear combination of the two. The choice of clustered DSFG model has no impact on the tSZ power and we adopt the power-law template as the baseline model.
} 
  \label{fig:1halo2halo}
\end{figure}

In addition to the choice of one template or the other, we extend the baseline model to include a second DSFG clustering component.
The two DSFG clustered amplitudes, corresponding to the power-law and the linear-theory templates, are allowed to vary independently.
The spectral index for all three (Poisson and two clustered) DSFG components are tied.
The two-dimensional likelihood surface for the amplitudes of each template is shown in Figure \ref{fig:1halo2halo}.
The data show a slight preference for the power-law template over the linear-theory template; however, we can not rule out the linear-theory model or a linear combination of the two with any confidence.
We can expect this constraint on DSFG clustering shape to improve with future data.

\subsection{tSZ-DSFG correlation}
\label{subsec:dusty-tsz-correl}

As discussed in \S\ref{subsec:tszdsfgcorrel}, there are reasons to expect that the signal from the tSZ and the clustered DSFGs may be correlated.  
In order to study that correlation, we extend the model to include the correlation coefficient of these components, $\gamma(\ell)$, detailed in \S\ref{subsec:tszdsfgcorrel}.
We assume the correlation coefficient is constant in $\ell$, thus reducing $\gamma(\ell)$ to $\gamma$.
We would expect an anticorrelation at frequencies below the SZ null; however, we do not require that the
correlation coefficient is negative.
Instead, we set a flat prior on the correlation $\gamma$, and allow it to range from -1 to 1.

We consider the impact of introducing correlations between the tSZ and DSFGs on two chains -- both allow the correlation factor to vary and one also allows the kSZ amplitude to vary.
In both chains, we assume the linear-theory clustered DSFG template, because its shape is similar to the S10 tSZ template.

\begin{figure}
\includegraphics[width=0.5\textwidth]{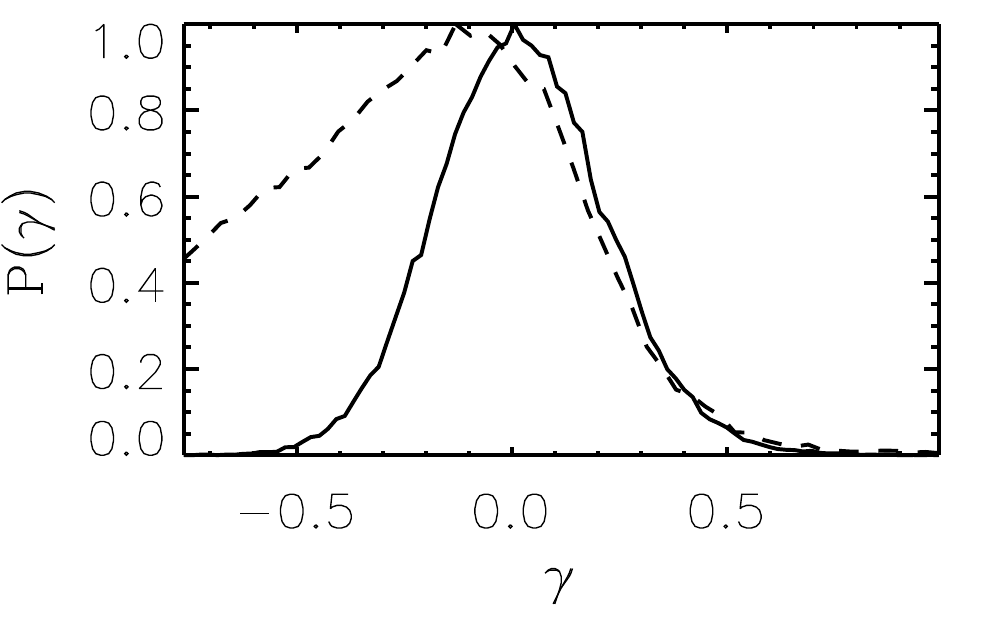}
  \caption[]{
The 1D likelihood curve for the tSZ-DSFG correlation factor, $\gamma$.
  The linear-theory clustered DSFG template and the S10 tSZ template are assumed.
The {\bf solid line} indicates the case where the kSZ amplitude is held fixed. 
The data favor no correlation albeit with significant uncertainty.
The {\bf dashed line} denotes the likelihood curve when the kSZ amplitude is allowed to vary.
The data remain consistent with zero correlation, but large negative correlations become possible. 
Essentially, the kSZ power increases to balance the lower predicted power due to anti-correlation.
Because the tSZ and kSZ amplitudes are highly correlated, additional kSZ power reduces the 
apparent tSZ power.
  }
\label{fig:szdsfgcorr}
\end{figure}

\begin{figure*}[ht]\centering
 
  \includegraphics[width=0.9\textwidth]{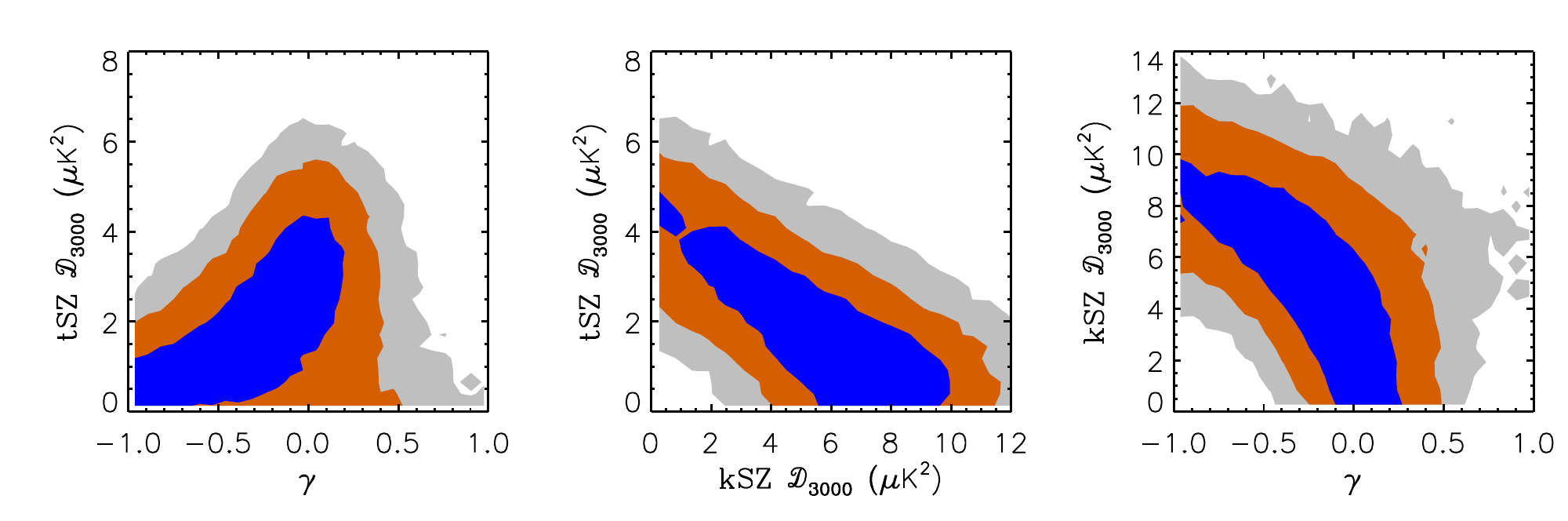}
  \caption[]{2D likelihood contours for tSZ-DSFG correlation chains in which kSZ is allowed to vary.  
    The 68\%, 95\% and 99.7\% likelihood contours are marked with blue, orange and gray respectively.
    From left to right: tSZ power vs. correlation coefficient $\gamma$; 
    tSZ power vs. kSZ power, marginalized over $\gamma$ (and foreground terms); and
    kSZ power vs. $\gamma$.
  } 
  \label{fig:like2dtszkszcorrelated}
\end{figure*}

The results for this model without the possibility of a tSZ-DSFG correlation have been presented 
in \S\ref{subsec:twohalo}.
The marginalized 1d likelihood function for the tSZ-DSFG correlation factor is shown in Figure \ref{fig:szdsfgcorr}.
In the fixed kSZ case, the data are consistent with zero correlation: $\gamma = 0.0 \pm 0.2$. 
When the kSZ power is allowed to vary, the data remain consistent with zero correlation,
however, large anti-correlations are also possible.
The 95\% CL upper limit on $D^{tSZ}_{3000}$ allowing for a tSZ-DSFG correlation is unchanged from the equivalent chain with the 
correlation coefficient set to zero; marginalizing over a non-zero tSZ-DSFG correlation factor 
does not allow additional tSZ power.
As Figure~\ref{fig:like2dtszkszcorrelated} shows, this is because kSZ power increases for larger negative
correlation in order to preserve the observed power in the $150\times220$ bandpowers.
Naively, one might expect the best fit tSZ power to increase when the possibility of a tSZ 
canceling correlation is taken into account;
however, in practice, a negative correlation leads to less tSZ power in the best fit model.
The anti-correlation reduces the predicted power in the $150\times220$ spectrum, and kSZ power 
is increased to compensate.  
The sum of kSZ and tSZ is well constrained by the data, so more kSZ power leads to less tSZ power.
When marginalizing over the tSZ-DSFG correlation, the $95\%$ CL on tSZ decreases slightly to $D^{tSZ}_{3000}<4.5\,\mu{\rm K}^2$, while the kSZ upper limit increases to $D^{kSZ}_{3000}<9.7\,\mu{\rm K}^2$.
From this, as well as the agreement between the multi-frequency fits and 
the DSFG-subtracted result presented in \S\ref{sec:l10s10comp},
we conclude that the addition of a tSZ-DSFG correlation cannot explain the 
discrepancy between the observed tSZ power and models that predict a significantly
larger tSZ contribution.

These results depend on the Poisson and clustered DSFG terms having similar spectral indices.
We expect the upper limits to be substantially weakened if the two spectral indices are allowed to separate, but note that the required spectral index separation would need to be much larger than the $\alpha_p-\alpha_c \sim$0.2 difference expected from models (H10).

\subsection{Comparing results from differenced spectra to multi-frequency fits}
\label{sec:l10s10comp}

The main result of this work is the set of multi-frequency bandpowers. However, we also follow the treatment in L10 and consider a linear combination of the two frequency maps designed to eliminate the foreground contribution from DSFGs.
Specifically, this is the power spectrum of the map $m$ constructed from the 150 and $220\,$GHz maps ($m_{150}$ and $m_{220}$ respectively) according to
\begin{equation}
 m = \frac{1}{1-x} (m_{150} - x \,m_{220}).
 \end{equation}

\begin{figure*}[ht]\centering
\includegraphics[width=0.9\textwidth]{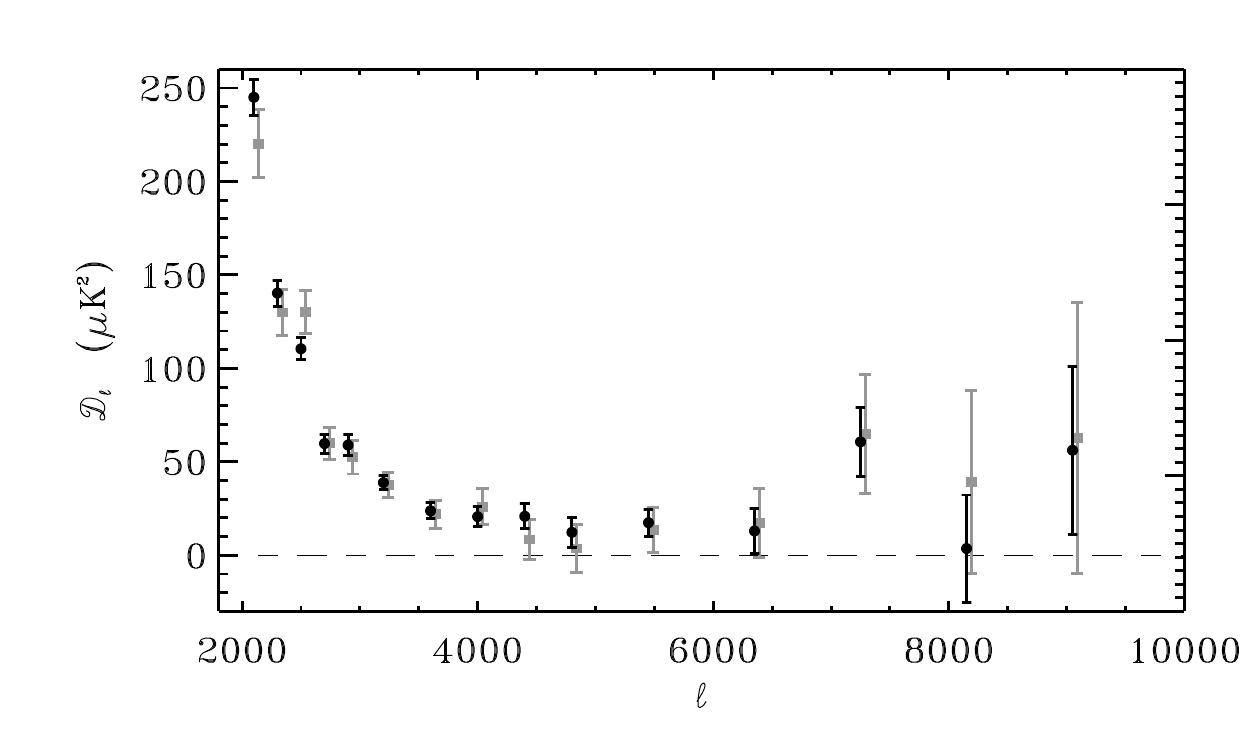}
  \caption[]{The SPT DSFG-subtracted bandpowers from this work ({\bf black circles}) versus the DSFG-subtracted bandpowers presented in L10 ({\bf gray squares}), an earlier reduction of one half of the 2008 data.
  The damping tail of the primary CMB anisotropy is apparent below $\ell = 3000$. 
  Above $\ell = 3000$, there is a clear excess due to secondary anisotropies and residual point sources.
  The two datasets show excellent consistency.
}
 
\label{fig:l10comp}
\end{figure*}

In L10, the subtraction coefficient $x$ was selected by minimizing the Poisson point source power in the resultant bandpowers. 
We repeat this analysis for the new bandpowers and find $x = 0.312 \pm 0.06$.
This is consistent with the value of $x = 0.325 \pm 0.08$ reported in L10.
For complete details on the analysis, we refer the reader to that work.
Given the consistency of these results, we have chosen to use $x = 0.325$ to create the `DSFG-subtracted' bandpowers presented in Table \ref{tab:bandpowersdif} to enable the direct comparison of the two data sets.
These two subtraction coefficients correspond to a spectral index for the DSFG population of $\alpha_p$ = 3.7 and 3.6 respectively, consistent with the $\alpha_p = 3.57 \pm 0.09$ determined in multi-frequency fits.
We compare the L10 bandpowers with those presented in this work in Figure \ref{fig:l10comp}, and find excellent agreement between them.
Note that the two data sets are not independent since L10 analyzed one of the two fields used in this work.

\begin{table*}[ht!]
\begin{center}
\caption{\label{tab:bandpowersdif} DSFG-subtracted bandpowers}
\small
\begin{tabular}{cc|cc}
\hline\hline
\rule[-2mm]{0mm}{6mm}
$\ell$ range&$\ell_{\rm eff}$ &$\hat{D}$ ($\mu{\rm K}^2$)& $\sigma$ ($\mu{\rm K}^2$) \\
\hline
2001 - 2200 & 2056 & 245.0 &   9.6  \\ 
2201 - 2400 & 2273 & 140.3 &   6.9  \\ 
2401 - 2600 & 2471 & 110.5 &   6.0  \\ 
2601 - 2800 & 2674 &  59.8 &   5.1  \\ 
2801 - 3000 & 2892 &  59.0 &   5.4  \\ 
3001 - 3400 & 3185 &  38.9 &   3.9  \\ 
3401 - 3800 & 3581 &  23.8 &   4.3  \\ 
3801 - 4200 & 3994 &  20.7 &   5.4  \\ 
4201 - 4600 & 4402 &  20.9 &   6.7  \\ 
4601 - 5000 & 4789 &  12.3 &   7.9  \\ 
5001 - 5900 & 5449 &  17.4 &   7.3  \\ 
5901 - 6800 & 6360 &  13.0 &  11.8  \\ 
6801 - 7700 & 7257 &  60.7 &  18.4  \\ 
7701 - 8600 & 8161 &   3.6 &  28.6  \\ 
8601 - 9500 & 9063 &  56.2 &  44.8  \\ 
 
\hline
\end{tabular}
\tablecomments{ Band multipole range and weighted value $\ell_{\rm eff}$, bandpower $\hat{D}$, 
and uncertainty $\sigma$ for the DSFG-subtracted auto-spectrum of the SPT fields. 
This is the power spectrum of a linear combination of the 150 and 220 GHz maps ($(m_{150} - 0.325 \,m_{220})/(1-0.325)$) designed to suppress the DSFG contribution.
The quoted uncertainties include instrumental noise and the Gaussian sample variance of the primary CMB and the point source foregrounds. 
The sample variance of the SZ effect, beam uncertainty, and calibration uncertainty are not included. 
Beam uncertainties are shown in Figure~\ref{fig:beam} and calibration uncertainties are quoted in \S\ref{sec:calibration}.
Point sources above $6.4\,$mJy at $150\,$GHz have been masked out in this analysis, eliminating the majority of radio source power. }
\normalsize
\end{center}
\end{table*}

This `DSFG-subtracted' power spectrum is interesting for two reasons.
First, it allows a direct comparison with the results of L10.
Second, in the DSFG-subtracted bandpowers, the residual correlated DSFG power will only be significant if the frequency spectrum of DSFGs in galaxy clusters is substantially different than the spectrum of DSFGs in the field.
Therefore, parameter fits to DSFG-subtracted bandpowers are independent of the assumed DSFG power spectrum shape or tSZ-DSFG correlation, and are a potentially interesting cross-check on the multi-frequency fits.
However, in practice, it is worth noting that these fits constrain a similar linear combination of tSZ and kSZ, roughly given by tSZ + 0.5 kSZ.
Both methods therefore result in a similarly robust detection of tSZ power independent of tSZ-DSFG correlation.

For this DSFG-subtracted case, we simplify the baseline model to include only one free parameter beyond $\Lambda$CDM: the tSZ power spectrum amplitude.
We also include a Poisson point source term with a strict prior, and a fixed kSZ power spectrum.
The S10 model is used for both kSZ and tSZ.
The Poisson prior is identical to the point source prior described in L10 with one exception:
the radio source amplitude prior has been tightened to match the 15\% prior used in this work instead of the 50\% log-normal prior used in L10.
However, tests with relaxed priors show the results are insensitive to this radio uncertainty.
Except for this small change in the radio prior, and the switch to WMAP7 data, these differenced spectrum fits are identical to the parameter fitting used in L10.

As shown in Figure \ref{fig:aszlike1d}, the resulting DSFG-subtracted tSZ amplitude agrees with both the L10 value and the multi-frequency result presented here.  We measure the tSZ power at 150 GHz to be $D_{3000}^{tSZ} = 3.5 \pm 1.2\,\mu {\rm K}^2$, with a fixed (S10 homogeneous) kSZ component. 
For the same template choices, L10 report 
$D_{3000}^{tSZ} = 3.2 \pm 1.6\,\mu {\rm K}^2$, and the multi-frequency fit to the baseline model finds $D^{tSZ}_{3000}=3.5 \pm 1.0\,\mu {\rm K}^2$. 
This close agreement demonstrates the consistency of the two methods of parameter fitting, and confirms the results presented in L10.

\section{Consequences of the measured \lowercase{t}SZ Signal}
\label{sec:sigma8}

\begin{figure}
\includegraphics[scale = 0.40]{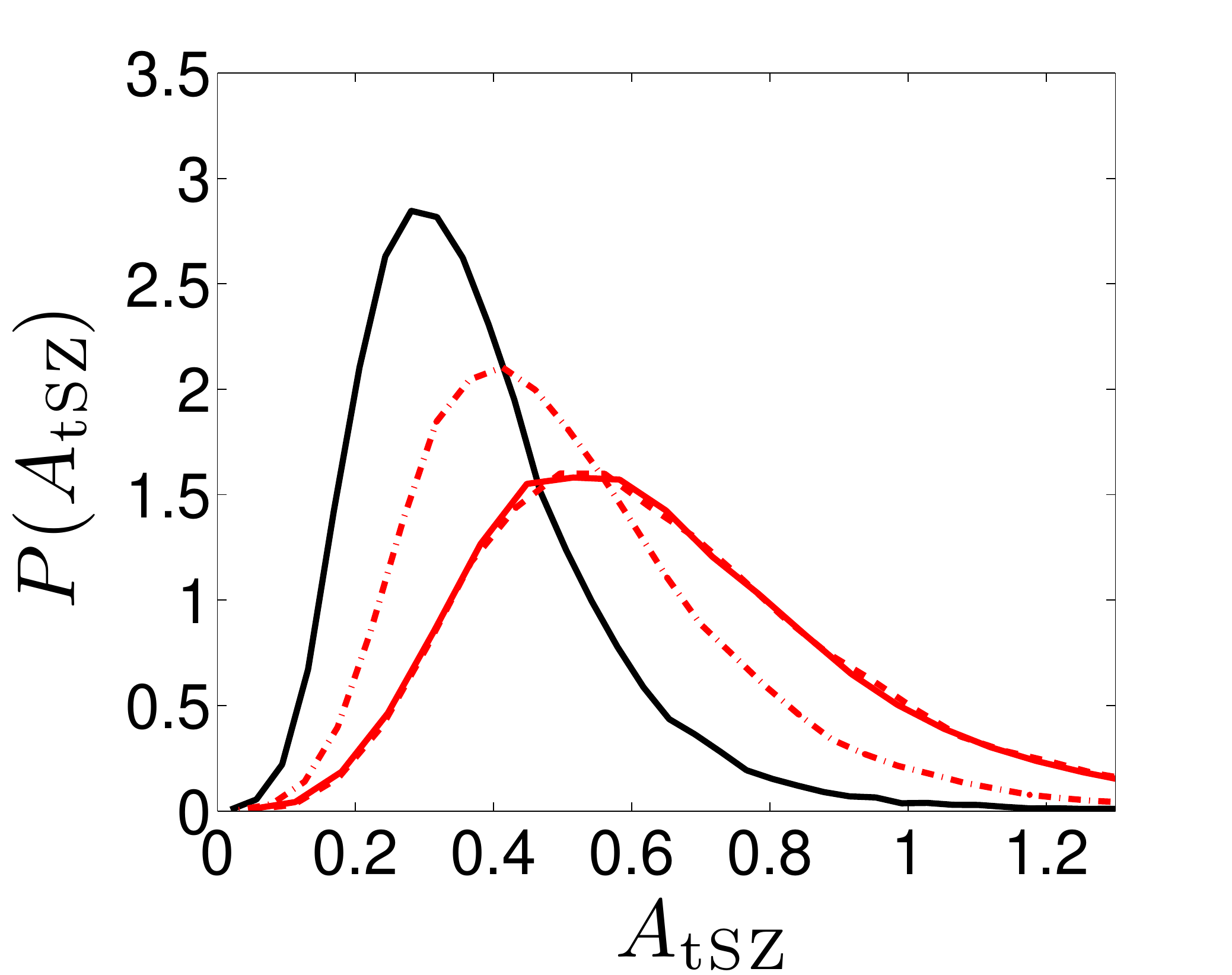}
\caption{Posterior distribution of the tSZ scaling parameter $\atszcosm$ (see
  Equation \ref{eq:atszcosm} and Table \ref{tab:tszconstraints}) for each of the tSZ templates. The {\bf black line} shows the S10 prediction. The {\bf red solid, dashed and dot-dashed lines} show the results for the Shaw, Trac, and Battaglia models, respectively. 
  As expected, the preferred value of $A_{\rm tSZ}$ increases for models which predict less tSZ power. 
  Note that a value of  $\atszcosm = 1$ would demonstrate consistency between the predicted and observed tSZ power.}
\label{fig:tsz_model_constraint}
\end{figure}

We now focus on comparing the improved measurements of the tSZ power
spectrum with the predictions of recent models and simulations. We
then combine measurements of the primary CMB power spectrum with the
tSZ signal to obtain joint constraints on $\sigma_8$.

\subsection{Thermal SZ amplitude}

L10 found the measured amplitude of the tSZ power spectrum to
be a factor of $0.42 \pm 0.21$ lower than that predicted by the
simulations of S10 (at their fiducial cosmology). In this analysis, having
doubled the survey area to 210 deg$^2$, we find that the amplitude
of the tSZ power spectrum at $\ell = 3000$ is $D_{3000}^{tSZ}=3.5 \pm 1.0\,\mu{\rm
  K}^2$ at $150\,$GHz. This is $0.47 \pm 0.13$ times the S10
template.

The discrepancy suggests that either the S10 model overestimates the tSZ power, or the assumed set of cosmological parameters may be wrong.  
To calculate the tSZ power spectrum, S10 combined a
semi-analytic model for intra-group and cluster gas  with a cosmological
N-body lightcone simulation charting the matter distribution out to
high redshift.  The gas model assumed by S10 was calibrated by
comparing it with X-ray observations of low-redshift ($z < 0.2$)
clusters. However, the majority of tSZ power is contributed by groups
or clusters at higher redshift \citep{komatsu02, trac10}.  
Hence, the S10 model may substantially overpredict
the contribution from higher redshift structures. Note that in 
\S\ref{subsec:dusty-tsz-correl}, we demonstrated that the
lower-than-predicted tSZ power is unlikely to be due to a
spatial correlation between DSFGs and the groups and clusters that
dominate the tSZ signal.

As described in \S\ref{sec:tszmodels}, in addition to the S10
template we investigate results obtained using the templates suggested
by \citet{trac10}, \citet{shaw10} and \citet{battaglia10} (the Trac, Shaw, and Battaglia models).  
These three alternative models all predict substantially less
power at $\ell = 3000$ than the S10 template (assuming the same
cosmological parameters). For each template we run a new chain
substituting the S10 template with one of the three others. The rest
of the baseline model remains as described in \S\ref{sec:model}.

Alternatively, the amplitude of the tSZ power spectrum is extremely sensitive to $\sigma_8$, scaling approximately as
$D_\ell^{tSZ} \propto \sigma_8^{7-9} (\Omega_b h)^2$. 
The exact
dependence on $\sigma_8$ depends on the relative contribution of groups and clusters 
as a function of mass and redshift to the power spectrum, and thus the
astrophysics of the intra-cluster medium \citep{komatsu02, shaw10, trac10}. 
All four of the templates that we consider in this work were determined assuming $\sigma_8 = 0.8$. 
The deficit of tSZ power measured in the SPT maps may therefore indicate a lower value of $\sigma_8$. 

To correctly compare the measured tSZ power spectrum amplitude with theoretical predictions, we must rescale the template power spectra so that they are consistent with the cosmological parameters at each point in the chains. 
Recall that the cosmological parameters are derived solely from the joint WMAP7+QUaD+ACBAR+SPT constraints on the primary CMB power spectrum and so are not directly influenced by the amplitude of the measured tSZ power.

We define a dimensionless tSZ scaling parameter, $\atszcosm$, where
\begin{equation}
A_{\rm tSZ}(\kappa_i) = \frac{D_{3000}^{tSZ}}{\Phi_{3000}^{tSZ}(\kappa_i)} \;.
\label{eq:atszcosm}
\end{equation}
$D_{3000}^{tSZ}$ is the measured tSZ power at $\ell = 3000$ and $\Phi_{3000}^{tSZ}(\kappa_i)$ is the template power spectrum normalized to  be consistent with the set of cosmological parameters `$\kappa_i$' at step $i$ in the chain. We obtain the preferred value of $\atszcosm$ by combining the measurement of $D_{3000}^{tSZ}$ with the cosmological parameter constraints provided by the primary CMB measurements.

To determine the cosmological scaling of each of the templates, we adopt the
analytic model described in \citet{shaw10}. This model uses the halo
mass function of \citet{tinker08} to calculate the abundance of dark
matter halos as a function of mass and redshift, combined with an
analytic prescription \citep[similar to that of][]{bode09} to
determine the SZ signal from groups and clusters. \citet{shaw10}
demonstrated that when the same assumptions are made about the stellar mass content and feedback processes in groups and clusters, their model almost exactly reproduces the S10 template. 
For the Trac and Battaglia models, we modify the astrophysical parameters of
the Shaw model so that it agrees with the other models at
their fiducial cosmology. 
Varying the input cosmological parameters
thus allows us to estimate the cosmological scaling of the tSZ power
spectrum for any given model.

As an indication of the relative cosmological scaling of each model at
$\ell = 3000$, in Table \ref{tab:cosmscaling} we give the dependence of $\Phi_{\rm tSZ}$ on  $\sigma_8$, $\Omega_b$, and $h$ around their fiducial
values. We assume that the cosmological scaling takes the form,
\begin{equation}
\Phi^{\rm tSZ}_{3000} \propto \left(\frac{\sigma_8}{0.8}\right)^\alpha\left(\frac{\Omega_b}{0.044}\right)^\beta\left(\frac{h}{0.71}\right)^\gamma \;.
\label{eq:cosmscale}
\end{equation}
Note that the values of $\alpha$ given in Table \ref{tab:cosmscaling}
for the S10 and Trac models agree well with those given in
\citet{trac10}. We find the templates depend only weakly on other cosmological parameters at this angular scale.

\begin{table}[ht!]
\begin{center}
\caption{\label{tab:cosmscaling} tSZ Cosmological Scaling}
\small
\begin{tabular}{c|c|c|c}
\hline\hline
\rule[-2mm]{0mm}{6mm}
model & $\alpha$ & $\beta$ & $\gamma$\\
\hline
Sehgal & 8.09 & 2.50 & 1.89 \\
Shaw & 8.34 & 2.81 & 1.73\\
Trac &  8.34 & 2.90 & 1.81 \\
Battaglia & 8.33  & 2.85 & 1.71\\
\hline
\end{tabular}
\normalsize
\tablecomments{$\alpha$, $\beta$, and $\gamma$ describe the cosmological scaling of the tSZ power spectrum for each template, as defined in Equation \ref{eq:cosmscale}.}
\end{center}
\end{table}

Figure \ref{fig:tsz_model_constraint} shows the posterior probability of
$\atszcosm$ obtained for the S10, Shaw, Trac, and Battaglia templates. 
The results are also given in Table
\ref{tab:tszconstraints}. 
Recall that a value of $\atszcosm = 1$ would
demonstrate that the model is consistent with the measured tSZ power,
while $\atszcosm$ less (greater) than 1 indicates that the model over-
(under-) estimates the tSZ power spectrum amplitude. 
All four
models display a non-Gaussian distribution with tails extending to
higher values of $\atszcosm$. 
Therefore, in Table \ref{tab:tszconstraints}, we give  both symmetric 68\% confidence limits for both $\atszcosm$ and $\ln ({\atszcosm})$ as well as the  asymmetric 95\% confidence limits (in brackets) for $\atszcosm$.

For the S10 template, the preferred value is $\atszcosm = 0.33\pm 0.15$. The measured level of tSZ power is thus significantly less than that predicted by the S10 template. When the kSZ amplitude is also allowed to vary, we find a slight decrease in $\atszcosm$ to $0.30 \pm 0.16$.  Note that the constraint on $\sigma_8$ obtained from the primary CMB power spectrum is $0.821 \pm 0.025$, which is greater than the fiducial value of $0.8$ assumed by all the templates.\footnote{We note that, following the upgrade to RECFAST v1.5 and in the WMAP likelihood code to v4.1, the WMAP7-alone preferred value of $\sigma_8$ is $0.811^{+0.030}_{-0.031}$ as given in Appendix B of \citet{larson10}.} The values of $\atszcosm$ are therefore less than what we would have obtained had we fixed the templates at their fiducial amplitudes.

We find
that the discrepancy remains, but is reduced, for the Shaw, Trac, and
Battaglia models. The Trac and Shaw templates provide the best match
to the measured amplitude with $\atszcosm = 0.59 \pm 0.25$.  It is interesting to note that the total amount of tSZ power measured ($D_{3000}^{tSZ}$ in Table \ref{tab:tszconstraints}) is essentially independent of the template used. This demonstrates that the differences in the {\it shapes} of the templates do not influence the results.

The templates that are most consistent with the observed SPT power at $\ell = 3000$ (when combined with cosmological constraints derived from the primary CMB power spectrum) are those that predict the lowest tSZ signal. The S10, Trac, and Shaw models adopt a similar approach for predicting
the tSZ power spectrum. Each assumes that intra-cluster gas resides in
hydrostatic equilibrium in the host dark matter halo, and include
simple prescriptions to account for star formation and AGN feedback. The
Trac and Shaw models -- which both predict significantly less tSZ
power -- differ from the S10 model in two principal
ways. 
First, following the observational constraints of \citet{giodini09}, they
assume a greater stellar mass fraction in groups and clusters. This reduces
the gas density and thus the magnitude of the SZ signal from these
structures. Second, the Trac and Shaw models include a significant
non-thermal contribution to the total gas pressure.
Hydrodynamical simulations consistently observe significant levels of nonthermal pressure in the intra-cluster medium due to bulk gas motions and turbulence \citep{evrard90, rasia04, kay04, dolag05,
  lau09}. In terms of the tSZ power spectrum, including a non-thermal
pressure component reduces the thermal pressure required for the ICM
to be in hydrostatic equilibrium, therefore reducing the predicted tSZ
signal.

The Battaglia template was produced using maps constructed from a
hydrodynamical simulation that included star formation and AGN
feedback. While direct comparisons with the model parameters of the
other three templates are difficult, \citet{battaglia10} found that,
in simulations in which AGN feedback was not included, the tSZ power
spectrum increased by 50\% at the angular scales probed by SPT. If we assume the tSZ template produced from their simulations without AGN feedback we find $\atszcosm = 0.27 \pm 0.10$ and thus a significant increase in the discrepency between the observed and predicted tSZ power. \citet{battaglia10} demonstrate that the inclusion of AGN feedback reduces the thermal gas pressure in the central regions of clusters by inhibiting the inflow of gas. This results in flatter radial gas pressure profiles and a suppression of tSZ power on angular scales $\ell > 2000$. 

\subsection{$\sigma_8$ Constraints}

\begin{table*}[ht!]
\begin{center}
\caption{\label{tab:tszconstraints} \lowercase{t}SZ Constraints}
\small
\begin{tabular}{c|c|c|c|c}
\hline\hline
\rule[-2mm]{0mm}{6mm}
model  &  $D^{tSZ}_{3000}$ ($\mu{\rm K^2}$) & $\atszcosm$ &  $\ln (\atszcosm)$ & $\sigma_8$\\
\hline
Sehgal & $3.5 \pm 1.0$ & $ 0.33 \pm 0.15 \,(^{+0.40}_{-0.23})$   & $-1.14 \pm 0.45$  & $0.771 \pm 0.013$ \\
Sehgal, free kSZ & $3.3 \pm 1.3$ &  $ 0.30 \pm 0.16 \,(^{+0.39}_{-0.30}) $  & $-1.27 \pm 0.47$  & $0.772 \pm 0.013$ \\
Shaw  & $3.6 \pm 1.0$ & $0.58 \pm 0.27  \,(^{+0.67}_{-0.40})$ & $-0.54 \pm 0.40$   & $0.799 \pm 0.014$ \\
Trac & $3.6 \pm 1.0$ & $0.59 \pm 0.25 \,(^{+0.72}_{-0.41})$   & $-0.53 \pm 0.45$  & $0.802 \pm 0.013$ \\
Battaglia & $3.5 \pm 0.9$ & $0.45 \pm 0.20 \,(^{+0.53}_{-0.31})$  & $-0.79 \pm 0.44$ &$ 0.786 \pm 0.014$\\
\hline
\end{tabular}
\tablecomments{
Measured tSZ power and inferred $\sigma_8$ constraints for different tSZ and kSZ model assumptions.
Note that the posterior distribution of $\atszcosm$ is  non-Gaussian (see Figure \ref{fig:tsz_model_constraint}). 
  To account for this we give both the 68\% and asymmetric 95\% (in brackets) confidence intervals. We also give the median and 68\% confidence interval for $\ln (\atszcosm)$. For $D^{tSZ}_{3000}$ and $\sigma_8$ we quote only the 68\% confidence interval.  
  The $\sigma_8$ constraints do not account for model uncertainty.
  All results assume the homogeneous reionization kSZ template given in S10. 
  In the `free kSZ' case the amplitude of this template is allowed to vary, otherwise it is held fixed at $2.05 \,\mu{\rm K}^2$.}  
\normalsize
\end{center}
\end{table*}

\begin{figure}
\includegraphics[scale = 0.40]{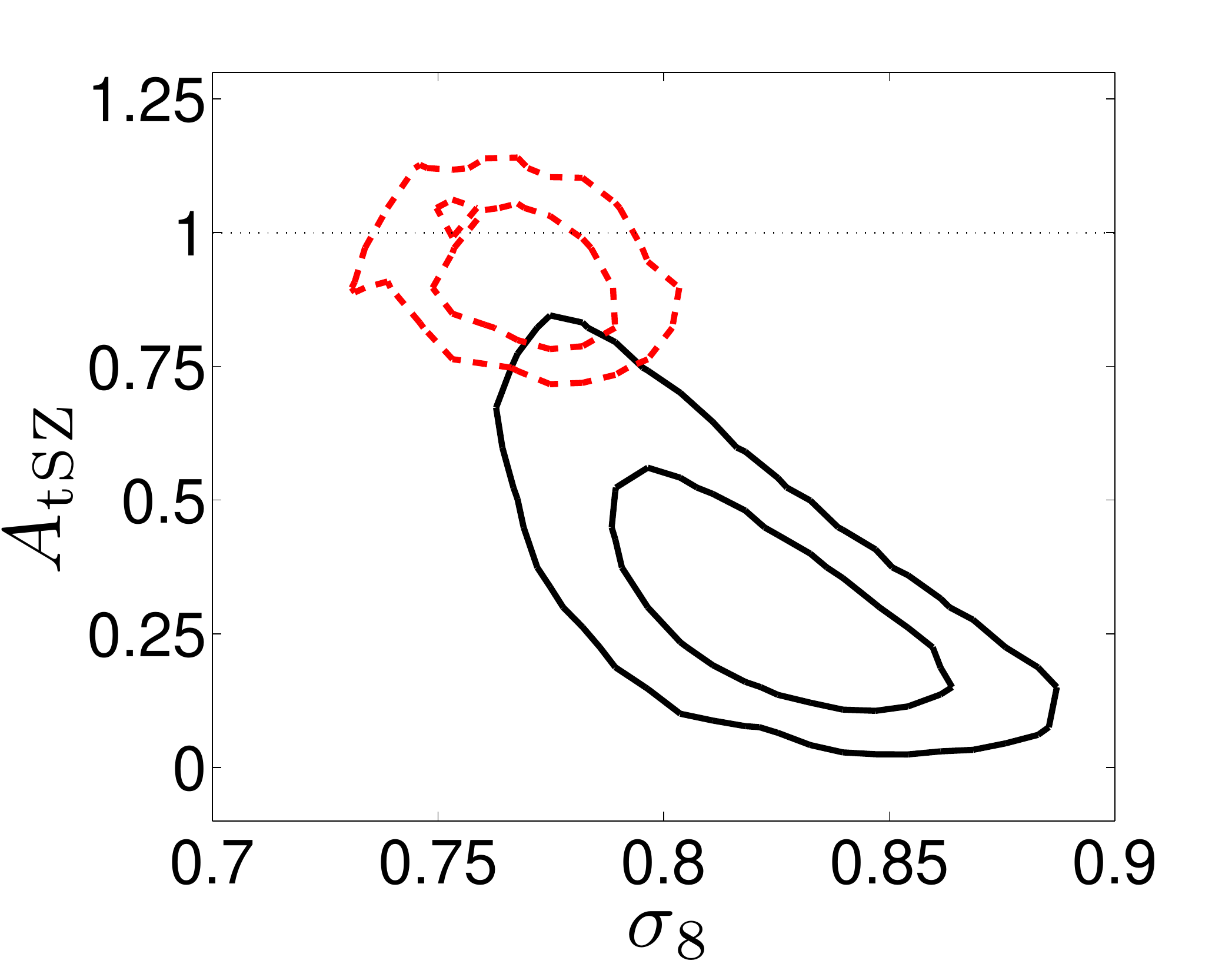}
\caption{2D likelihood contours at 68\% and 95\%
  confidence for $\sigma_8$ vs. the tSZ scaling parameter $\atszcosm$
  (see Eqn. \ref{eq:atszcosm}). 
  {\bf Red, dashed contours} show results having
  resampled the chains applying a lognormal prior to $\atszcosm$
  centered at $\ln (\atszcosm) = 0$ and of width
  $\sigma_{\ln (\atszcosm)} = 0.085$.
  This prior is chosen to reflect the expected sample variance for the observed sky area (see text). 
  {\bf Black contours} show
  the results without this prior. All results are for the baseline
  model (S10 tSZ model and fixed homogeneous kSZ). }
\label{fig:asz_sigma8_baseline}
\end{figure}

\begin{figure}
\includegraphics[scale = 0.40]{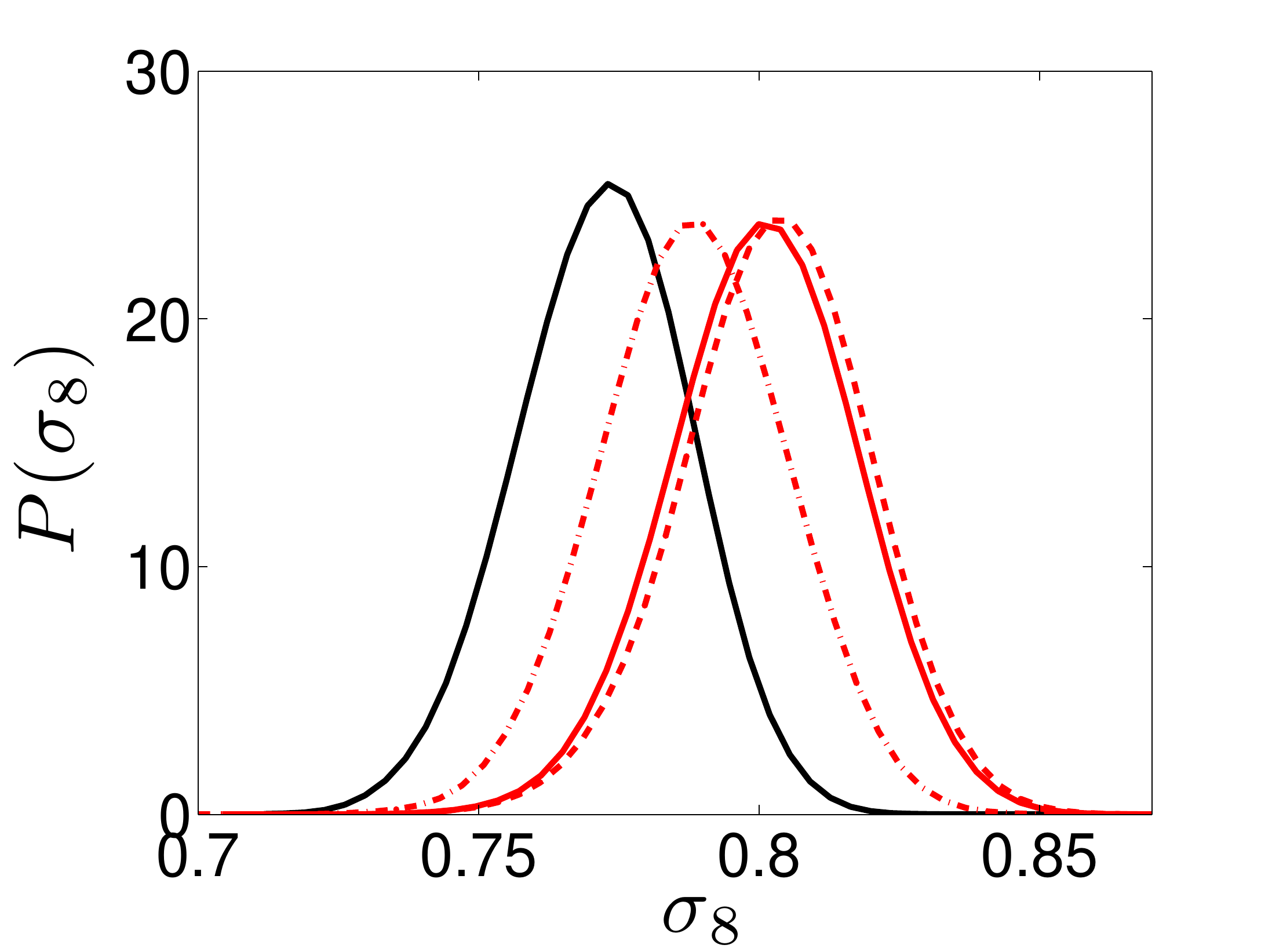}
\caption{$\sigma_8$ constraints for the S10 ({\bf black line}), Shaw ({\bf red,
  solid line}), Trac ({\bf red, dashed line}) and Battaglia ({\bf red, dot-dashed line}) models.
  No model uncertainty has been included.}
\label{fig:sigma8_models}
\end{figure}

To obtain joint constraints on $\sigma_8$ from both the primary CMB
anisotropies and the tSZ amplitude we must determine the probability of
observing the measured value of $\atszcosm$  for a given set of
cosmological parameters. In the absence of cosmic variance -- and assuming the predicted tSZ template to be perfectly accurate -- the
constraint on $\sigma_8$ would be given by its distribution at
$\atszcosm=1$ for each of the models. However, in practice, the
expected sample variance of the tSZ signal is non-negligible. Based on
the hydrodynamical simulations of \citet{shaw09}, L10 assumed the sample variance at
$\ell = 3000$ to be a lognormal distribution of mean $0$ and width $\sigma_{\ln (\atszcosm)} = 0.12$.  The survey
area analyzed here is twice that of L10 and thus we reduce the width of this distribution by a factor of $\sqrt{2}$. For each of the four
tSZ templates, we construct new chains to include this prior by
importance sampling. We address the issue of theoretical uncertainty in the tSZ templates later in this section.

The black contours in Figure \ref{fig:asz_sigma8_baseline} show the 68\%
and 95\% confidence limits in the $\sigma_8 - \atszcosm$ plane for the
S10 model before applying the prior on $\atszcosm$.  In this case, the
constraint on $\sigma_8$ is derived entirely from the primary CMB power
and is not influenced by the measured tSZ signal. The combined WMAP7+ACBAR+QUaD+SPT constraints on the primary CMB yield  $\sigma_8 =
0.821 \pm 0.025$. The shapes of the contours reflect the dependence of
$\atszcosm$ on $\sigma_8$; higher values of $\sigma_8$ predict more
tSZ power and require a lower tSZ scaling parameter, whereas lower values
of $\sigma_8$ suggest a tSZ amplitude closer to that predicted by the
S10 template.

The dashed red contours in Figure \ref{fig:asz_sigma8_baseline} represent the
joint 68\% and 95\% confidence limits on $\sigma_8$ and $\atszcosm$
for the S10 template having resampled the chain to include the prior
on $\atszcosm$. The 1D $\sigma_8$ constraints for all four templates are shown in Figure \ref{fig:sigma8_models} and Table
\ref{tab:tszconstraints}.  For the S10 template the preferred value of
$\sigma_8$ is $0.771 \pm 0.013$.\footnote{During the analysis for this work we discovered that the joint primary CMB-tSZ constraint on $\sigma_8$ in \cite{lueker10} erroneously assumed the S10 model was created using $\sigma_8= 0.77$ instead of the correct value of $\sigma_8=0.8$. The constraints on $\sigma_8$ should therefore have been $0.767 \pm 0.018$ \citep[assuming no theory uncertainty, see Section 6.4 in][]{lueker10} and $0.778 \pm 0.022$ (assuming a 50\% theory uncertainty).} The
Trac, Shaw, and Battaglia models all prefer higher values of
$\sigma_8$. The Trac template -- the most consistent with the measured
tSZ amplitude -- gives $\sigma_8 = 0.802 \pm 0.013$. This represents nearly a factor of two
improvement in the statistical constraint on $\sigma_8$ compared to that obtained from the primary CMB alone.

It is clear from Table \ref{tab:tszconstraints} that the four templates produce a spread in the preferred value of $\sigma_8$  that is larger than the statistical uncertainty on any of the individual results. This spread reflects the range in amplitude of the tSZ templates (at a fixed value of $\sigma_8$) due to the differences in the thermal pressure profiles of groups and clusters in the underlying models and simulations. The dominant uncertainty on the tSZ-derived $\sigma_8$ constraint is thus the {\it astrophysical} uncertainty on the predicted tSZ amplitude rather than the statistical error on the measurement of $D_{3000}^{\rm tSZ}$. Accounting for this uncertainty will clearly degrade the constraint on $\sigma_8$.

To produce a constraint on $\sigma_8$ that accounts for the range in amplitude of the tSZ templates we must add an additional theory uncertainty to the prior on $\atszcosm$. Assuming the Trac template -- which is the most consistent with the measured tSZ signal -- we find that a 50\% theory uncertainty on $\ln ({\atszcosm})$ is sufficient to encompass the other three templates. As the sample variance is insignificant compared to the theory uncertainty, the new prior on $\atszcosm$ is effectively a lognormal distribution of width $\sigma_{\ln (\atszcosm)} = 0.5$. With this additional uncertainty, the revised constraint on $\sigma_8$ for the Trac template is $0.812 \pm 0.022$. The astrophysical uncertainty clearly degrades the constraint significantly -- it now represents only a small reduction of the statistical uncertainty of the primary CMB-only result.

It is clear that, while the amplitude of the tSZ power spectrum is extremely sensitive to the value of $\sigma_8$, the constraints that can obtained on this parameter are limited by the accuracy with which we can predict the tSZ power spectrum amplitude for a given cosmological model. The principal difficulty in producing precise predictions is that groups and clusters spanning a wide range of mass and redshift contribute to the power spectrum at the angular scales being probed by SPT. For instance, \citet{trac10} find that 50\% of the power in their nonthermal20 template is due to objects at $z \geq 0.65$. Similarly, they find that  50\% of the power comes from objects of mass less than $2\times 10^{14} \hmsun$. To date, both these regimes have been poorly studied by targeted X-ray and SZ observations. 
It is therefore difficult to test current models and simulations of the tSZ power 
spectrum sufficiently well to produce precise predictions. 
X-ray and SZ observations of groups and high redshift clusters will provide constraints on models and simulations that should improve estimates of the tSZ power spectrum and thus allow tighter constraints on $\sigma_8$ from measurements of this signal \citep{sun10}.
Alternatively, better constraints on cosmological parameters and improved measurements of the tSZ power 
spectrum will continue to shed new light on the relevant astrophysics in low mass and high redshift 
clusters.

\section{Conclusions}
\label{sec:conclusions}

We have presented the CMB temperature anisotropy power spectrum from the
complete $210\,{\rm deg}^2$ of sky observed by SPT in the 2008 season.
Bandpowers from observations at $150$ and $220\,$GHz are shown in
Table~\ref{tab:bandpowerssinglefreq}.  The bandpowers correspond to
angular scales, $\ell>2000$, where secondary anisotropies and
foreground contributions are significant.
We also present bandpowers that have been spectrally
differenced to remove the power contributed by dusty sources, 
following the method described in the first SPT power spectrum release (L10).  

We perform multi-frequency model fits to the $\ell > 2000$ bandpowers to constrain secondary CMB anisotropies and foreground signals.
The minimal extension to the six-parameter lensed $\Lambda$CDM primary CMB spectrum that we consider consists of four free parameters: the amplitudes of the tSZ spectrum, Poisson and clustered DSFG terms, and the spectral index of the DSFG sources.
The addition of these four parameters improves the likelihood of the best-fit model by $\Delta {\rm ln} \mathcal{L} = -3980$.
This baseline model also includes a fixed kSZ contribution and a tight prior on radio sources.
We explore the effect of modifying this model by allowing additional model parameters to vary.
In particular, we free the amplitude of the kSZ and the clustered DSFG spectral index.
However, we do not find that these additional free parameters improve the 
quality of fits to the data.

With these data, the amplitudes of the two SZ components are largely
 degenerate. The data primarily constrain a linear combination of tSZ and kSZ,
 with the result $D^{tSZ}_{3000} + 0.5\,D^{kSZ}_{3000} = 4.5\pm 1.0 \,\mu{\rm
 K}^2$. For the baseline model with the kSZ power spectrum fixed to that
  expected for homogeneous reionization, we constrain the amplitude of the tSZ
  power at $\ell = 3000$ to be $D^{tSZ}_{3000}= 3.5\pm 1.0\, \mu{\rm K}^2$. We
  test the dependence of the measured power on the specific tSZ model power  
  spectrum and find that with the present data, the result is insensitive to 
  the model shape for several recently published models. We also find that the
  measured tSZ power is insensitive to the modeling assumptions for the DSFG
  and radio foregrounds. We derive upper limits on secondary anisotropy power
  when the kSZ amplitude is allowed to vary. At 95\% CL, we find
  $D^{tSZ}_{3000} < 5.3\,\mu{\rm K}^2$ and $D^{kSZ}_{3000} < 6.5\,\mu{\rm
  K}^2$. The corresponding limit on the total SZ power is $D^{tSZ}_{3000} +
  0.5\,D^{kSZ}_{3000} < 6.2\,\mu{\rm K}^2$ at the 95\% CL.

We compare the measured amplitude of the tSZ power spectrum with the
predictions of recent models and simulations. 
L10 found less tSZ power than the  \citet[S10]{sehgal10} model predicts;
the results in this paper increase the significance of that claim.
We find that, for the
cosmological parameters preferred by the primary CMB power spectrum,
the simulations of S10 overpredict the tSZ power spectrum amplitude, with an
observed-to-predicted ratio of $\atszcosm = 0.33 \pm 0.15$. The more recent models of \citet{shaw10} and \citet{trac10}
are less discrepant, but still overestimate the tSZ signal. For example,
the `nonthermal20' model of \citet{trac10} gives $\atszcosm = 0.59 \pm 0.25$. All three of these models assume that intra-cluster
gas resides in hydrostatic equilibrium in the host dark matter
gravitational potential, and include prescriptions to account for
star formation and AGN feedback. The \citet{trac10} and
\citet{shaw10} models also include a significant non-thermal
contribution to the total gas pressure, which results in a lower
tSZ signal. For the cosmological hydrodynamical simulations including star formation and AGN feedback
of \citet{battaglia10}, we find $\atszcosm = 0.45 \pm 0.20$.  If instead we adopt the tSZ template derived from their simulations that do not include star formation and AGN feedback we find $\atszcosm = 0.27 \pm 0.10$ and thus a larger discrepancy between the measured and predicted signal.

We allow for a correlation between the tSZ and DSFG components in the model to investigate whether such a correlation could be responsible
for the measurement of lower than expected tSZ power.
With the kSZ amplitude fixed, marginalizing over the tSZ-DSFG correlation coefficient has no 
significant effect on the resulting tSZ power. 
When the kSZ power and tSZ-DSFG correlation are both free to vary, the fits actually prefer a larger kSZ power and slightly reduced tSZ power.
We also note that if tSZ-DSFG correlation did significantly affect the measured tSZ power, one would expect a difference between the result of the multi-frequency fits that assume no tSZ-DSFG correlation and fits to the differenced bandpowers, which should be largely free of DSFG signal.
Instead, both analyses find the same tSZ power.
We therefore conclude that correlation between the tSZ and DSFG components cannot be 
used to reconcile the measured tSZ power with model predictions, 
at least when operating under the reasonable assumption that 
the Poisson and clustered DSFG spectral indices are similar.

We also investigate combining the measured tSZ amplitude with CMB data
to obtain joint constraints on $\sigma_8$. We find that, if we assume that the template predictions are perfectly accurate, the
statistical uncertainty is reduced by as much as a factor of two compared to the
primary CMB-only constraints. 
The resulting values of $\sigma_8$ range from $0.771 \pm 0.013$ for the
S10 model to $0.802 \pm 0.013$ for the \citet{trac10}
predictions. 
If we assume that the predicted amplitude is uncertain at 50\% in order to account for the range 
of model predictions, the tSZ measurement adds little additional information, and 
the constraint degrades to $\sigma_8=0.812 \pm 0.022$. Current modeling uncertainties prevent the use of the tSZ power spectrum as a tool for precision cosmology, 
however, precise measurements of the tSZ power spectrum can provide new information about cluster 
physics in the poorly studied regimes of low mass and high redshift.

In addition to the constraints on SZ power and $\sigma_8$, these data allow us to place constraints on foreground components.
The measured bandpowers are consistent with a simple foreground model consisting of clustered DSFGs with a single spectral index and an unclustered distribution of radio sources (with a strong prior from radio source count models).
Using this model, 
we find the spectral index of the DSFGs is $\alpha=3.58 \pm 0.09$. 
The Poisson amplitude is $D^p_{3000}=7.4 \pm 0.6\,\mu{\rm K}^2$ and the clustered amplitude is $D^c_{3000}=6.1 \pm 0.8\,\mu{\rm K}^2$. 
We explore a number of variations on this simple DSFG model and, in general, find
these variations have little impact on the SZ constraints.
By comparing fits using alternate models of the clustered DSFG power spectrum, we find 
the data show a slight preference for the power-law model over the linear-theory model used in H10.

We find a Poisson DSFG spectral index of $\alpha_{p}=3.58 \pm 0.09$.
Although the Poisson spectral index is well-constrained and fairly robust to changes in the model, 
the amplitude of the Poisson power changes by $4\,\sigma$ as we change the shape of the
clustered template.  
The Poisson power estimates are therefore dominated by this modeling uncertainty,
and any analysis of these results needs to take that uncertainty into account.
The constraint on the clustering spectral index has improved from  $3.8 \pm 1.3$ reported in H10 to $\alpha_{c}=3.47 \pm 0.34$.  
This spectral index is sensitive to the shape of the clustered DSFG template:
with the power-law template, we find $\alpha_{c}=3.79 \pm 0.37$.
In either case, the data are consistent with identical indices for both the Poisson and clustering DSFG contributions to the power spectrum.

In 2009, the SPT receiver was refurbished to add sensitivity at $95\,$GHz.
Since then, we have mapped an additional $1300\,{\rm deg}^2$ to depths of 42, 18, and $85\,\mu$K-arcmin at 95, 150, and $220\,$GHz respectively.
These data will break the tSZ/kSZ degeneracy present in the current data and 
facilitate the separation of the SZ components.
Improved measurements of the tSZ power spectrum will provide new insight to 
the cluster astrophysics relevant in the poorly understood regimes of low 
mass and high redshift. 
We also expect these data to provide new constraints on the kSZ power spectrum, 
which have the potential to illuminate the epoch of reionization.

\begin{acknowledgments}

The South Pole Telescope is supported by the National Science
Foundation through grants ANT-0638937 and ANT-0130612.  Partial
support is also provided by the NSF Physics Frontier Center grant
PHY-0114422 to the Kavli Institute of Cosmological Physics at the
University of Chicago, the Kavli Foundation and the Gordon and Betty
Moore Foundation.  
The McGill group acknowledges funding from the National   
Sciences and Engineering Research Council of Canada, 
Canada Research Chairs program, and 
the Canadian Institute for Advanced Research. 
M. Dobbs acknowledges support from an Alfred P. Sloan Research Fellowship.
L. Shaw acknowledges the support of Yale University and NSF grant AST-1009811.  
M. Millea and L. Knox acknowledge the support of NSF grant 0709498.  
This research used resources of the National Energy Research Scientific Computing Center, which is supported by the Office of Science of the U.S. Department of Energy under Contract No. DE-AC02-05CH11231. 
Some of the results in this paper have been derived using the HEALPix \citep{gorski05} package. 
We acknowledge the use of the Legacy Archive for Microwave Background Data Analysis (LAMBDA). 
Support for LAMBDA is provided by the NASA Office of Space Science.
\end{acknowledgments}

\clearpage

\bibliography{../../BIBTEX/spt.bib}

\end{document}